\documentclass[fleqn,usenatbib,onecolumn]{mnras}

\usepackage{newtxtext,newtxmath}

\usepackage[T1]{fontenc}
\usepackage{ae,aecompl}

\usepackage{graphicx}	
\usepackage{amsmath}	
\usepackage{amssymb}	
\usepackage{epsfig}

\date{Accepted 2018 July 29. Received 2018 July 13; in original form 2018 February 8}

\pubyear{2017}

\title[FRB event rate counts II]{FRB event rate counts II --- fluence, redshift and dispersion measure distributions}
\author[Macquart \& Ekers]{J.-P. Macquart,$^{1,2}$\thanks{\href{mailto:J.Macquart@curtin.edu.au}{J.Macquart@curtin.edu.au}} and R.D. Ekers,$^{3,1}$  \\
$^{1}$International Center for Radio Astronomy Research, Curtin University, GPO Box U1987, Perth, WA 6845, Australia\\
$^{2}$ARC Centre of Excellence for All-Sky Astrophysics (CAASTRO)\\
$^{3}$CSIRO Astronomy and Space Science (CASS), P.O. Box 76, Epping, NSW 1710, Australia\\
}

\begin{document}
\label{firstpage}
\pagerange{\pageref{firstpage}--\pageref{lastpage}}
\maketitle

\begin{abstract}
We examine how the various observable statistical properties of the FRB population relate back to their fundamental physical properties in a model independent manner. 
We analyse the flux density and fluence distributions of Fast Radio Bursts (FRBs) as a tool to investigate their luminosity distance distribution and the evolution of their prevalence throughout cosmic history.   We examine in detail particular scenarios in which the burst population follows some power of the cosmic star formation rate. FRBs present an important additional measurable over source counts of existing cosmological populations, namely the dispersion measure.  Based on the known redshift of FRB121102 (the repeater) we expect at least 50\% of the dispersion measure to be attributable to the inter-galactic medium and hence it can be used as a proxy for distance. We develop the framework to interpret the dispersion measure distribution, and investigate how the effect of Helium reionization in the intergalactic medium is evident in this distribution. Examination of existing data suggests that the FRB luminosity function is flatter than a critical slope, making FRBs easily detectable to large distances; in this regime the reduction in flux density with distance is outweighed by the increase in the number of bright bursts within the search volume.  
\end{abstract}


\section{Introduction}
The progenitors of the Fast Radio Burst (FRB) population are presently a subject of intense speculation. The lack of any definitive model for an FRB motivates an approach, adopted here, to examine how the FRB observables of flux density, fluence and dispersion measure (DM), are related to the intrinsic properties of an FRB in a model independent manner.  This requires a few broad assumptions. The DMs of these millisecond-duration events place them outside the galaxy \citep{Thorntonetal2013}, and the localisation of FRB121102 to a galaxy at $z = 0.19$ \citep{Chatterjeeetal2017,Tendulkaretal2017} shows that at least 50 percent of the DM for this FRB is attributable to the intergalactic medium.  Thus it is reasonable to proceed on the assumption that the larger DMs are ascribed primarily to the intergalactic medium \citep[see, e.g.,][]{Lorimeretal2007,Thorntonetal2013} and that, while the DM contributions from the host galaxies will increase the scatter, they will not destroy the DM-distance relationship.  
However almost nothing is known about the distribution of the population with distance.  This is another important clue in unravelling their origin, since it would reveal how the abundance of FRBs has evolved throughout cosmic time.  One obvious means of attacking this problem at present is through an understanding of the FRB event rate counts. The distribution of FRB fluences offers a means of decrypting the identity of the progenitor population because it is coupled to the distributions of the luminosities and event distances, and hence to the evolutionary history of the progenitors. 

In a companion paper \citep[][hereafter Paper I]{MacquartEkers18}, we describe the venerable history and the proven usefulness of source count statistics in the analysis of other astrophysical populations, such as quasars and gamma-ray bursts, and recount how they were employed to assess the distribution of these sources over cosmological distances. Paper I outlines in detail the motivation for investigating FRB source counts in particular.  It discusses the treatment of various biases inherent to the current FRB sample, and it derives the present observational constraints on the event rate distribution.  Application of a maximum likelihood technique to the Parkes data indicates that the index of the integral rate counts distribution,  ${\cal R}(>F_\nu) \propto F_\nu^{\beta}$ as a function of limiting fluence, $F_\nu$, is steep, with $\beta = -2.6_{-1.3}^{+0.7}$ at $F_\nu > 2\,$Jy\,ms; this constraint invites interpretation in the context of the evolution of the FRB population.  

The purpose of the present paper is to elucidate how the various observable statistical properties of the FRB population relate back to their fundamental physical properties.  Though the theory of source counts statistics is well understood in the context of radio galaxies, active galactic nuclei and gamma ray bursts, FRBs add a new dimension to the problem because each detection is accompanied by its dispersion measure (DM).  The ability to measure the DM distribution for any extragalactic population represents a new and powerful diagnostic of its properties and, potentially, its distance and evolution. 

Analysis of the DM distribution of FRBs would be a potent cosmological tool providing that a considerable portion of each FRB dispersion measure is attributable to the intergalactic medium (IGM).  This question provides the motivation to undertake the present study: the predicted characteristics of the DM distribution can be used to investigate the validity of this hypothesis and, if proven, would constitute a means of probing the distribution in detail.  Such quantities are particularly useful in the present era, when FRB localisations are currently scarce \citep[the only localisation being that of][] {Tendulkaretal2017} and their distances are largely unknown.  However, even when the burst redshifts are known it will be necessary to understand how the measured underlying redshift and DM distribution relates to the detection parameters of a given survey, especially through its sensitivity and spectral resolution.  



In this paper we have investigated the event rate distributions of the FRB population in terms of the observables; flux density, fluence and dispersion measure as determined by the underlying luminosity function and its redshift dependence. We place particular emphasis on properties involving the fluence.  This is motivated by the fact that the observed flux density of an impulsive radio burst is affected by both the detector temporal resolution (as discussed in Paper I), and by temporal smearing of the pulse due to multipath propagation.  Temporal smearing is known to be an important effect for FRBs \citep[e.g.][]{Lorimeretal2007,Thorntonetal2013,MacquartKoay2015}, but its effects are not well understood, there being no clear relation between the dispersion measure of an FRB and its scattering timescale, thus rendering its incorporation into the source counts formalism problematic.    However, a treatment of the fluence distribution obviates the need to account for finite detector resolution, and thus the distribution of burst durations relative to it.  The time-integrated pulse energy is also invariant to the scattering timescale for a statistically homogeneous scattering medium (but may deviate from this if the assumption of statistical homogeneity is violated on scales from which the scattered emission is received).  If more sophisticated detection criteria (e.g. matched filtering) are used for the survey this will affect the completeness fluence, but one would expect an analysis of its effects to form a part of the survey completeness analysis, rather than an intrinsic component of the source counts theory.


The paper is partitioned as follows. In \S\ref{sec:Definition} we introduce the event rate formalism for a flux-density or fluence limited survey.  In \S\ref{sec:Distribution} we apply this theory to derive the behaviour of these distributions for various broadly-generic FRB evolutionary scenarios.  In \S\ref{sec:RateandDMcompare} we introduce the formalism to derive the DM distribution of the population.  In \S\ref{sec:Discussion} we illustrate the application of this formalism using a comparison with the DM distribution of of published events and discuss the implications of our findings.  Our conclusions are presented in \S\ref{sec:Conclusions}.

\section{Event counts formalism} \label{sec:Definition}

In this section we briefly review the formalism that links the flux density and fluence distribution of a population of events (or sources) to its luminosity function and evolutionary history \citep[see][for an insightful discussion of the general problem]{vonHoerner1973}.  This distribution depends on the number of sources seen per luminosity\footnote{We give a list of symbols used throughout this text in Table \ref{tab:Symbols}, noting in particular that our usage of the terms fluence and luminosity follows conventions in the FRB and pulsar fields, but which is otherwise non-standard.} in the range $L_\nu$ to $L_\nu + dL_\nu$ over the volume element $dV$.  In a Euclidean universe, the volume of a thin shell at distance $r$ is $dV = 4 \pi r^2 dr d\Omega$, so that the number of events seen over this luminosity and radius range $dr$ over a solid angle $d\Omega$ is
\begin{eqnarray}
n(L_\nu,r) dL_\nu dr d\Omega = r^2 dr \Phi_L(L_\nu;r) dL_\nu d\Omega, \label{nLBasicEq}
\end{eqnarray}
where $\Phi(L_\nu;r) dL_\nu$ is the luminosity distribution function per unit volume, with units of comoving density (hereafter termed the ``luminosity function'' for brevity).  Evolution in the population is incorporated in the dependence of the luminosity function on distance or, equivalently, redshift, $z$.

\begin{table*}
\caption{A list of the most important symbols used throughout the text, their names, clarification of their meaning (if required), and their dimensions.
$\dagger$ The use of the term ``fluence'' in the FRB field is non-standard, in that it refers to the energy per unit area per unit bandwidth (instead of energy per unit area).  Accordingly, for consistency with this definition we adopt the same non-standard usage and refer to $L_\nu$ and $E_\nu$ as the ``luminosity'' and ``energy'' respectively .  We note that in the future wideband receivers may detect FRBs across the entire extent of their radio emission, in which case they will directly measure the energy per unit area.     
} 
\begin{tabular}{llll}
\hline
Symbol &
Name & Remarks &
Representative Units  \\
 \hline
 $D_H$ & Hubble distance & $c/H_0$ & Mpc \\
 $D_C$ & comoving distance & & Mpc \\
 $D_A$ & angular diameter distance & & Mpc \\
 $D_L$ & luminosity distance & & Mpc \\ \hline
 $F_\nu$ & fluence$^\dagger$ & energy spectral density per unit area & Jy\,ms \\
 $S_\nu$ & flux density & power spectral density per unit area & Jy \\
 $L_\nu$ & luminosity$^\dagger$ (spectral density) & total power output spectral density, $dL/d\nu$ & Jy\,m$^2$ \\
 $E_\nu$ & energy$^\dagger$ (spectral density) & total energy output spectral density, $dE/d\nu$ & Jy\,ms\,m$^2$ \\ 
 $\alpha$ & spectral index of the flux density or fluence & defined by $S_\nu \propto \nu^{-\alpha}$ or $F_\nu \propto \nu^{-\alpha}$ & dimensionless \\
\hline
${\cal R}$ & total event rate & & events\,s$^{-1}$\,sr$^{-1}$ \\
 $ {\cal R}(>F_\nu)$ & integral rate counts distribution & cumulative event rate above threshold fluence, $F_\nu$ & events\,s$^{-1}$\,sr$^{-1}$ \\
$\beta$ & source counts index  & index of cumulative fluence counts, ${\cal R} \propto F_\nu^\beta$  & dimensionless \\ \smallskip
$\frac{d{\cal R}}{dS_\nu}$ & differential (flux density) source counts & differential event rate wrt flux density &  events\,s$^{-1}$\,(Jy)$^{-1}$\,sr$^{-1}$ \\ \smallskip
$\frac{d{\cal R}}{dF_\nu}$ & differential (fluence) source counts & differential event rate wrt fluence & events\,s$^{-1}$\,(Jy\,ms)$^{-1}$\,sr$^{-1}$ \\
$R_{\rm obs}(S_\nu,z)$ & (differential) event rate in observer's frame & differential event rate wrt flux density and redshift & events\,s$^{-1}$\,(Jy)$^{-1}$\,sr$^{-1}$ \\
    & & ``obs'' refers to $z$-corrected rate in observer's frame & \\
$R_{\rm obs}(F_\nu,z)$ &  (differential) event rate in observer's frame  & differential event rate wrt fluence and redshift & events\,s$^{-1}$\,(Jy\,ms)$^{-1}$\,sr$^{-1}$ 
\\
    & & ``obs'' refers to $z$-corrected rate in observer's frame & \\
$\Phi_L(L_\nu ; z)$  & luminosity function & number per unit comoving volume per $L_\nu$ at $z$  & events\,m$^{-3}$\,(Jy\,m$^2$)$^{-1}$ \\
$\Phi_E(E_\nu ; z)$ & energy function & number  per unit comoving volume per $E_\nu$ at $z$  & events\,m$^{-3}$\,(Jy\,ms\,m$^2$)$^{-1}$ \\
$\Theta_L(L_\nu ; z)$  &  event rate luminosity function & $\Theta_L(L_\nu;z) \equiv d\Phi_L(L_\nu;z)/dt_{\rm emit}$ & events\,s$^{-1}$\,m$^{-3}$\,(Jy\,m$^2$)$^{-1}$ \\ 
& & event rate per comoving volume per $L_\nu$ at $z$ & \\
$\Theta_E(E_\nu ; z)$ &  event rate energy function & $\Theta_E(E_\nu;z) \equiv d\Phi_E(E_\nu;z)/dt_{\rm emit}$  & events\,s$^{-1}$\,m$^{-3}$\,(Jy\,ms\,m$^2$)$^{-1}$ \\ 
& & event rate per comoving volume per $E_\nu$ at $z$ & \\
\hline
$L_{\rm min,max}$ & minimum and maximum $L_\nu$ & lower and upper luminosity density bounds of $\Theta_L$ & Jy\,m$^2$ \\
$E_{\rm min,max}$ & minimum and maximum $E_\nu$  & lower and upper energy density bounds of $\Theta_E$ & Jy\,ms\,m$^2$ \\
$\gamma$ & luminosity/energy function index & index of the (power law) functions of $\Theta_L$ or $\Theta_E$   & dimensionless \\
$\psi(z)$ & event rate per comoving volume at $z$ & & events\,s$^{-1}$\,Mpc$^{-3}$ \\ 
\hline
\end{tabular} \label{tab:Symbols}
\end{table*}

\subsection{Counts over cosmological distances} \label{sec:countsgeneral}

We consider an event counts model for a population whose emission emanates at cosmological distances in which the abundance may evolve significantly over time.  When the population of interest is distributed over a sufficiently large distance range (or range of lookback times), the volume element $dV$ must be generalised to take into account the spacetime geometry of the Universe, for which \citep[e.g.][]{Hogg1999}
\begin{eqnarray}
dV = D_{\rm H} \frac{(1+z)^2 D_{\rm A}^2}{E(z)} d\Omega dz,
\end{eqnarray}
where the term $D_{\rm H} dz/E(z)$ is related to $dr$ and $(1+z)^2 D_{\rm A}^2$ to the $r^2$ term in the Euclidean treatment.
The quantity $D_{\rm H} = c/H_0$ is the Hubble distance and $E(z) = \sqrt{\Omega_m (1+z)^3 + \Omega_k (1+z)^2 + \Omega_\Lambda}$ relates the present Hubble constant to its value at a redshift $z$: $H(z) = H_0 E(z)$.  The quantity $D_A$ is the angular diameter distance, $D_A = D_M/(1+z)$, where for the $\Omega_k = 0$ Universe assumed in this treatment\footnote{We also adopt values of $\Omega_\Lambda=0.7$, $\Omega_m = 0.3$, and $H_0 = 70\,$km\,s$^{-1}$\,Mpc$^{-1}$.}, the transverse comoving distance, $D_{\rm M}$, is related to the line-of-sight comoving distance, $D_C$ by
\begin{eqnarray}
D_{\rm M} = D_{\rm C} = D_{\rm H} \int_0^z \frac{dz'}{E(z')}.
\end{eqnarray}

Thus the number of objects in a redshift interval $dz$ and luminosity interval $dL_\nu$ is
\begin{eqnarray}
n(L_\nu,z) dL_\nu d\Omega dz  = \left( D_{\rm H} \frac{(1+z)^2 D_{\rm A}^2}{E(z)} d\Omega dz \right) \Phi_L(L_\nu;z) dL_\nu. \label{BasicNdistn}
\end{eqnarray}

The conversion between the event counts in terms of luminosity in the frame of the emitting source and flux density in the frame of the observer employs the relation
\begin{eqnarray}
S_\nu = (1+z) \frac{L_{(1+z)\nu}}{L_\nu} \frac{L_\nu}{4 \pi D_{\rm L}^2}, \label{S-Lrelation}
\end{eqnarray}
where $D_{\rm L}$ is the luminosity distance, defined by $D_{\rm L} = (1+z) D_{\rm M} = (1+z)^2 D_{\rm A}.$
The ratio $L_{(1+z)\nu}/L_\nu$ represents the $k$-correction due to the fact that the event emitted its radiation in a different band to that in which it is observed.  We writing the spectrum as 
\begin{eqnarray}
S_\nu \propto \nu^{-\alpha},
\end{eqnarray} 
so the $k$-correction is $(1+z)^{-\alpha}$.  


An additional consideration relevant to the detection of impulsive events such as FRBs is that the fundamental quantity of interest is not the total number of detections, but rather the detection rate.  This requires a correction to account for the fact that the event rate in the frame of the observer, $R_{\rm obs}(S_\nu,z) \equiv dn(S_\nu,z) /dt_{\rm obs}$ is dilated by a factor $(1+z)$ relative to that at the redshift of the emission: $R_{\rm obs}(S_\nu,z) = (dn(S_\nu,z) /dt_{\rm emit})/(1+z)$ (Throughout, we use the subscript ``obs'' to refer to the correction of the detection rate to the observer frame.)  Applying the operation $d/dt_{\rm obs}$ to both sides of eq.(\ref{BasicNdistn}), writing the differential luminosity in terms of flux (or fluence) density, and writing the angular diameter distance and luminosity distance in terms of comoving distance, we obtain 

\begin{eqnarray}
R_{\rm obs}(S_\nu,z) dS_\nu d\Omega dz &=& 4 \pi D_{\rm H}^5 \frac{(1+z)^{1+\alpha}}{E(z)} \left( \frac{D_{\rm M}}{D_{\rm H}} \right)^4 \frac{\Theta_L(L_\nu;z)}{1+z} dS_\nu d\Omega dz, \quad \hbox{where } L_\nu =\frac{4 \pi  D_{\rm L}^2}{(1+z)^{1-\alpha}} S_\nu, \label{SourceRate} 
\end{eqnarray} \label{RateEqns}
where we identify $\Theta_L(L_\nu;z) \equiv d\Phi_L(L_\nu;z)/dt_{\rm emit}$ as the event rate per luminosity per comoving volume from bursts at redshift $z$ in the frame of the emission (hereafter, for brevity, the event rate luminosity function), and use the fact that $d\Phi_L(L_\nu;z)/dt_{\rm obs} = [d\Phi_L(L_\nu;z)/dt_{\rm emit}]/(1+z)$.

\subsection{Fluence statistics}
Using the relationship between the burst fluence and the luminosity in the burst frame \citep[][]{MaraniNemiroff1996}
\begin{eqnarray}
F_\nu = \Delta t_e (1+z)^2 \frac{L_{(1+z)\nu}}{L_\nu} \frac{L_\nu}{4 \pi D_{\rm L}^2}, \label{F-Lrelation}
\end{eqnarray}
where $\Delta t_e$ is the burst duration in the frame of the emission, we can recast eqns.(\ref{nLBasicEq}) \& (\ref{BasicNdistn}) in terms of the fluence to derive
\begin{eqnarray}
R_{\rm obs} (F_\nu,z) dF_\nu d\Omega dz &=& 4 \pi D_{\rm H}^5 \frac{(1+z)^{\alpha}}{E(z)} \left( \frac{D_{\rm M}}{D_{\rm H}} \right)^4 \frac{\Theta_L(L_\nu;z)}{1+z} \frac{dF_\nu}{\Delta t_e} d\Omega dz, \quad \hbox{where } L_\nu =\frac{4 \pi  D_{\rm L}^2}{ \Delta t_e (1+z)^{2-\alpha}} F_\nu, \label{SourceRateF}
\end{eqnarray}

An alternate approach to the derivation of the fluence statistics involves working directly with the total energy of the event per unit bandwidth, $E_\nu \equiv \int L_\nu(t) dt$. In the foregoing formalism we have chosen to specify the rate statistics in terms of the underlying luminosity function, $\Phi_L(L_\nu;z)$.   Analogous to the definition of $\Phi_L$, we define the spectral energy density distribution of events (hereafter the ``energy function'') between $E_\nu$ and $E_\nu+ dE_\nu$ within the volume element $dV$ as $\Phi_E(E_\nu;z)$.  The equivalent rate relation for the fluence counts is
\begin{eqnarray}
R_{\rm obs} (F_\nu,z) dF_\nu d\Omega dz &=& 4 \pi D_{\rm H}^5 \frac{(1+z)^{\alpha}}{E(z)} \left( \frac{D_{\rm M}}{D_{\rm H}} \right)^4 \frac{\Theta_E(E_\nu;z)}{1+z} dF_\nu d\Omega dz, \quad \hbox{where } E_\nu =\frac{4 \pi  D_{\rm L}^2}{ (1+z)^{2-\alpha}} F_\nu, \label{SourceRateF2}
\end{eqnarray}
and where $\Theta_E(E_\nu;z) \equiv d \Phi_E(E_\nu;z)/dt_{\rm emit}$.

Since the flux density and fluence differ in their relation to luminosity by one power of $1+z$, there is a simple relation between the flux density and fluence statistics.  For a set of bursts of fixed rest-frame duration $\Delta t_e$, the differential fluence distribution is derived directly from the flux density distribution by making the replacements $\alpha \rightarrow \alpha -1$ and $dS_\nu \rightarrow d F_\nu/\Delta t_e$ in eq.(\ref{SourceRate}).   If instead we formulate the FRB statistics in terms of the spectral energy density distribution, $\Phi_E$, we can derive eq.(\ref{SourceRateF2}) directly from eq.(\ref{SourceRate}) using only the simple replacement: $\alpha \rightarrow \alpha -1$.   

Equations (\ref{SourceRate}) \& (\ref{SourceRateF2}) form the basis of our investigation into FRB counts.  The total event rate, which we shall label ${\cal R}$, is obtained by integrating $R_{\rm obs}(S_\nu,z)$ (or $R_{\rm obs}(F_\nu,z)$), the observed event rate per flux density (or fluence) per redshift, over all $z$ and $S_\nu$ (or $F_\nu$).  The primary quantities of interest are the derivatives of the event rate with respect to $S_\nu$ and $F_\nu$, which we label as $d{\cal R}/dS_\nu \equiv \int R_{\rm obs}(S_\nu,z) dz$ and $d{\cal R}/dF_\nu \equiv \int  R_{\rm obs}(F_\nu,z) dz$ respectively.   We make use of these basic relations throughout the remainder of this treatment.  

Finally, in the limit $z \ll1$ we remark that eqns.(\ref{SourceRate}) \& (\ref{SourceRateF2}) reduce to a particularly simple form for the flux density and fluence counts (von Hoerner 1973):
\begin{eqnarray}
R_{\rm obs} (S_\nu,z) 
 \, d\Omega \, dz dS_\nu = 4 \pi D_H^5 \, 
  \Theta_L(L_\nu;z) 
  d\Omega\, z^4 dz 
 dS_\nu,   \quad \hbox{where } 
   L_\nu =4 \pi  (z D_H)^2 dS_\nu . \label{lowzCounts}
\end{eqnarray}
The equivalent expression for fluence is obtained by substituting $S_\nu \rightarrow F_\nu$, $L_\nu \rightarrow E_\nu$ and $\Theta_L \rightarrow \Theta_E$.  In the limit $z \ll 1$, the $k$-correction is negligible, as are time dilation effects and the distinction between the angular diameter distance and luminosity distance.

Alternately, one can derive eq.(\ref{lowzCounts}) directly from eq.(\ref{nLBasicEq}) by transforming $r$ to flux density, $S_\nu = L/4 \pi r^2$, to obtain the flux density counts as a function of luminosity \citep[see][]{vonHoerner1973},
\begin{eqnarray}
n(L_\nu,S_\nu) dL_\nu dS_\nu d \Omega  =  4 \pi \left( \frac{L_\nu}{4 \pi S_\nu} \right)^{5/2} \frac{ \Phi_L(L_\nu;r)}{L_\nu} dL_\nu d\Omega. \label{EuclideanSoln}
\end{eqnarray}
One obtains eq.(\ref{lowzCounts}) upon further substitution of $L_\nu$ with $z$.  This is a statement of the well-known fact that the differential source counts for a Euclidean distribution is proportional to $S_\nu^{-5/2}$.

\section{The flux density and fluence distribution} \label{sec:Distribution}

In this section we first apply the foregoing formalism to the examine the event counts in the simple cases in which the distribution is confined to non-cosmological distances to elucidate a fundamental point about the source counts for nearby populations.  We then briefly summarise the statistics of a population of standard candles and standard ``batteries'' (i.e. events of fixed $E_\nu$) as a demonstration of the theory, despite being of unlikely relevance to the FRB population. The main results of this section are presented in \S\ref{sec:powerlawtreatment}, in which we summarise the well-known properties of the source counts for a power-law luminosity function, and investigate several specific formulations for the redshift evolution of the population that may apply to FRBs.  

A particular aspect of interest in this section relates to the finding in Paper I that the integral rate distribution appears to be a steep function of fluence, with index $d{\cal R}/dF_\nu \propto {F_{\nu}}^{\beta-1}$, with $\beta= -2.6_{-1.3}^{+0.7}$ \citep[see also][]{Bhandarietal17}.  This provides a motivation to explore how the assumption of a steep rate count distribution restricts the range of FRB progenitor models.

\subsection{Non-cosmological populations}


A well-known but noteworthy result is that any nearby extragalactic population (i.e.\,a non-cosmological one) exhibits counts that follow the Euclidiean source count distribution $dR_{\rm obs}/dS_\nu \propto S_\nu^{-5/2}$ and, equivalently, $dR_{\rm obs}/dF_\nu \propto F_\nu^{-5/2}$.  This result follows directly from eq.\,(\ref{EuclideanSoln}) for a nearby population irrespective of the luminosity function.   

It is useful to revisit the assumptions underpinning this result in view of its importance in the interpretation of several results pertaining to the more general treatment that follows in \S\ref{sec:powerlawtreatment}.  We explicitly use the formalism introduced above, applied to a population whose luminosity (or energy) follows a power law distribution but including a cut-off in luminosity which restricts the observable population to the local Universe at high flux densities.  We write the event rate luminosity function as
\begin{eqnarray}
\Theta_L (L_\nu;z) \equiv \frac{R_0}{A} L_\nu^{-\gamma} \label{PhiPower},
\end{eqnarray}  
where it is assumed that the abundance of events embodied in the redshift evolution of $\Theta(L_\nu;z)$ evolves slowly so that its dependence on $z$ is negligible over the small redshift range $0 < z \ll 1$. The luminosity function is taken to extend over the range $L_{\rm min} < L_\nu < L_{\rm max}$ where $A= [L_{\rm min}^{1-\gamma} - L_{\rm max}^{1-\gamma}]/(\gamma-1)$ is chosen so that $R_0$ is identified as the event rate per comoving volume\footnote{The quantity $\int_{L_{\rm min}}^{L_{\rm max}} \Theta (L_\nu) dL_\nu$ gives the rate density of sources per comoving volume.}.  
Integrating the expression derived for the count rate in the $z \ll 1$ limit, eq.(\ref{lowzCounts}), yields the differential source rate distribution,
\begin{eqnarray}
\frac{d {\cal R}}{dS_\nu}  
&=& \frac{ R_0 (\gamma-1) }{8 \pi^{3/2} (5 - 2 \gamma)} \frac{ L_{\rm max}^{5/2-\gamma}  - L_{\rm min}^{5/2-\gamma} }{L_{\rm min}^{1-\gamma} - L_{\rm max}^{1-\gamma} } S_\nu^{-5/2}, \qquad \gamma \neq 5/2. \label{EuclideanDistn}
\end{eqnarray}
The equivalent result, $d{\cal R}/dF_\nu \propto F_\nu^{-5/2}$, applies if one instead assumes a power law distribution in burst energies, $\Theta_E \propto E_\nu^{-\gamma}$, or if one simply substitutes $S_\nu \rightarrow F_\nu/\Delta t_e$ on the right hand side of eq.(\ref{EuclideanDistn}) for events of constant duration $\Delta t_e$.

The luminosity index $\gamma=5/2$ is a critical point at which the amplitude of the source counts distribution changes character from one dominated by the most luminous events (a shallow luminosity function, $\gamma < 5/2$) to the least luminous (a steep distribution, $\gamma > 5/2$).  For shallow luminosity functions, the redshift evolution of the event rate density can be neglected provided that the redshift of the brightest or most energetic detectable event,
\begin{eqnarray}
z_{\rm max} = \sqrt{\frac{H_0^2 L_{\rm max}}{4 \pi c^2 S_\nu}}, \quad \hbox{ or} \quad z_{\rm max} = \sqrt{\frac{H_0^2 E_{\rm max}}{4 \pi c^2 F_\nu}}, \label{zmax}
\end{eqnarray} 
is much less than 1.  This condition applies, for instance, to any sufficiently insensitive survey whose detection threshold, $S_\nu$ (or $F_\nu$), is so large that it is only possible to detect even the most luminous (most energetic) events in the nearby Universe.  Thus, for a cosmological population with a finite maximum luminosity there is always a sufficiently large flux density $S_\nu$ (or fluence, $F_\nu$) such that the redshift range of observed events is bounded to be so small, with $z_{\rm max} \ll 1$, that any evolution of $\Theta$ over this small range of observable cosmic time is negligible.  Above this critical flux density the source count distribution follows the Euclidean index of $-5/2$ for any population\footnote{This is valid up to a sufficiently high flux density at which the detection volume shrinks until the exact spatial distribution of events in the nearby Universe becomes important (i.e.\,\,if the events are tied to galaxies and the discrete distribution of host galaxies within the detection volume becomes important).}.  For steep luminosity functions the redshift cutoff, $z_{\rm max}$, is much lower than that defined by eq.\,(\ref{zmax}).
We emphasise that even with a sharp cut-off in luminosity it is unphysical for the event distribution to exhibit a sharp cutoff at high flux densities.

This result has important consequences for the high flux density tail of the event rate distribution.   In particular, it could be that diffractive scintillation boosting or lensing by plasma structures plays a part in the interpretation of FRB source counts \citep{MacquartJohnston2015,Cordesetal17}.  The probability of seeing an enhancement greater than a factor of $\mu$ over the mean flux density due to diffractive scintillation is $e^{-\mu}$ in the ideal case in which the decorrelation bandwidth is larger than the observing bandwidth.  

However, it is unlikely that diffractive scintillation is pushing a significant fraction of the bursts we detect over some detectability threshold.  In any source counts distribution which is intrinsically power law in nature, the boosting provided by diffractive scintillation can enhance the event rate, but the source counts distribution remains a power law of the same slope \citep[e.g.][]{MacquartJohnston2015}.  One may in principle obtain detections of much fainter events that have been boosted by scintillation, but since the exponential boosting probability distribution drops much faster than any power law distribution, any bursts one detects at a given flux density are on average only modestly enhanced.  Put another way, if one detects several FRBs at any given flux density this implies there must be a large number of such bursts with roughly that unboosted flux density.

Another possibility is that the source counts distribution has an abrupt cutoff at some limiting high flux density.  In that case, diffractive scintillation can draw a number of events to higher flux densities, resulting in an exponential tail in the counts distribution.  However, the foregoing results show that it is highly implausible to have a hard cutoff in a flux density (or fluence) distribution.  Once the horizon of observable events moves sufficiently nearby that spacetime curvature becomes negligible, the source counts always decline as $S_\nu^{-5/2}$, no matter what the luminosity function. One can, in principle, circumvent this argument if the bursts have a contrived nonuniform distance distribution (e.g. the luminosity function $\Phi_L(L;r)$ has a strange dependence on $r$, maybe such that all bursts are located on a thin shell), but this also appears implausible. Thus, to the extent that the high flux density source counts distribution always reverts to a power law, the statistics of the bursts should not become dominated by the exponential tail associated with diffractive scintillation.  

Alternate propagation-related models, particularly those associated with caustics produced by plasma lenses \citep[e.g.][]{Cordesetal17}, would lead to a different distribution of enhancements and possibly different conclusions.  However, the main point of this argument remains that the importance of enhancements due to propagation effects is constrained in a statistical sense by the source counts distribution.  Although large enhancements due to propagation effects are possible, their diminishing likelihood with magnification then dictates how far, in an average sense, the measured burst properties are likely to deviate from their intrinsic values.  

To illustrate this point quantitatively, consider the differential source counts distribution at some fluence $F_0$ resulting from an un-magnified (``intrinsic'') (differential) counts distribution $dR(F)/dF = K F^{\beta-1}$, with fluences extending from $F_{\rm min}$ to infinity, whose events are magnified by some process whose amplification probability distribution is $p_a$ \cite[see, e.g., the discussion relating to eq.(4) of][]{MacquartJohnston2015}:
\begin{eqnarray}
\frac{dR (F_0)}{dF} = K \int_{F_{\rm min}}^\infty F^{\beta-1} p_a \left(\frac{F_0}{F} \right) \frac{dF}{F}. 
\end{eqnarray}
It is instructive to recast this integral in terms of the magnification $\mu = F_0/F$, which yields
\begin{eqnarray}
\frac{dR (F_0)}{dF} = \frac{K}{F_0} \int_{0}^{F_0/F_{\rm min}} \left( \frac{F_0}{\mu} \right)^{\beta} p_a \left( \mu \right) d \mu, 
\end{eqnarray}
from which we deduce from the integrand that the condition for unmagnified events to dominate the differential rate at $F_0$ over those magnified by a factor $\mu$ is set by the approximate inequality
\begin{eqnarray}
p_a (1) \gtrsim \mu^{-\beta} p_a(\mu).
\end{eqnarray}
The magnification range in a given sample of events is bounded if the asymptotic decrease of $p_a(\mu)$ at large $\mu$ is steeper than $\mu^{\beta}$, specifically if the decline in the intrinsic integral source counts distribution is shallower than the decline in the probability of high-$\mu$ events. Then we may define a characteristic maximum magnification from the interplay between the magnification probability and the source counts slope, $\mu_0 \approx [p_a(1)/p_a(\mu_0)]^{-1/\beta}$.  For instance, for characteristic magnifications would be $\approx 5$ and $\approx 2$ respectively for $\beta=-2.6$ and $\beta=-1.5$ (Euclidean counts) if the magnification distribution scaled as $e^{-\mu}$.

\subsection{Standard candles}

Several treatments of FRB properties make use of the simplifying, albeit improbable (and even unphysical), assumption that FRBs radiate as standard candles \citep[e.g.][]{Lorimeretal2013}. The luminosity function is likely instead to be broad on the basis that that beaming effects are likely important and that the emission is necessarily coherent by virtue of its $\gg 10^{30}\,$K brightness temperature (c.f. the pulsar luminosity distribution).  The coherent emission mechanism will depend on the micro-physics which is almost certainly variable from source to source and for a repeating FRB from pulse to pulse.
Nevertheless, we briefly discuss the standard candle solution because the one-to-one mapping between flux density and redshift in this model elucidates the effects of cosmology on the rate distribution, without the additional complications introduced by a broad luminosity function.

A standard candle possesses an event rate luminosity function of the form,
\begin{eqnarray}
\Theta_L (L_\nu;z) = R_0  \psi(z) \delta (L_\nu-L_0), \label{StandCandleLum}
\end{eqnarray} 
where $\psi(z)$ describes the redshift evolution of the comoving rate density. If $\psi(z)$ is normalized so that $\psi(z) \rightarrow 1$ in the limit $z \rightarrow 0$, then $R_0$ would be interpreted as the event rate volume density at the present epoch, $z=0$.
The rate distribution is, from eq.\,(\ref{SourceRate}),
\begin{eqnarray}
\frac{d {\cal R}}{dS_\nu} &=& 4 \pi R_0 \, D_{\rm H}^5 \int_0^z \delta \left[g(z)\right] F(z) dz, \quad 
\hbox{where } F(z) = \psi(z) \frac{(1+z)^{\alpha}}{E(z)} \left[ \int_0^z \frac{dz'}{E(z')} \right]^4, \nonumber \\ 
&\null& \qquad \qquad \qquad \qquad \qquad \hbox{ and } g(z) = \frac{4 \pi  D_{\rm L}^2 S_\nu}{(1+z)^{1-\alpha}}  -L_0. \label{StandardC}
\end{eqnarray}
The integral over the Dirac delta function yields
\begin{eqnarray}
\frac{d {\cal R}}{dS_\nu}  =  4 \pi R_0 \, D_{\rm H}^5  \frac{F(z_i)}{|g'(z_i)|},
\end{eqnarray}
where $z_i$ is the root of $g(z)$.
%

The associated redshift distribution of standard candles detectable in a survey that reaches some minimum detectable flux density, $S_0$,  is obtained using eq.\,(\ref{SourceRate}):
\begin{eqnarray}
\frac{d {\cal R}}{dz} &=& \int_{S_0}^\infty R_{\rm obs}(S_\nu,z) \,dS_\nu \nonumber \\
&=& \begin{cases} 
R_0 \psi(z) \frac{D_H D_M^4}{D_L^2} \frac{1+z}{E(z)}, & z < z_c \\
0, & \hbox{otherwise}
\end{cases}  
\end{eqnarray}
where the cutoff redshift, $z_c$, is the solution to 
\begin{eqnarray}
S_0 = \frac{(1+z_c)^{1-\alpha} L_0}{4 \pi D_L(z_c)^2}, \label{zcDefn}
\end{eqnarray}
and where the luminosity distance is written here as an explicit function of redshift.

The standard candle event rate distribution is plotted in Figure \ref{fig:StandardCandle} for the assumption of a constant event rate per comoving volume, $\psi(z)=1$, and the associated redshift distribution is shown in Figure \ref{fig:StandardCandleZ}.  
The generic behaviour of the flux density distribution is interpretable in terms of the product of the redshift distribution, and the mapping between flux density and redshift: $d{\cal R}/dS_\nu = (d{\cal R}/dz) (dz/d S_\nu)$. 
The flattening observed in the flux density distribution for $S_\nu \lesssim 1$ relates directly to the turnover in the redshift distribution at $z \gtrsim 1.4$, since for the particular value of $L_0$ chosen here, a $z=1$ event has $S_\nu =1$.
(The turnover observed in the redshift distribution at $z \approx 1.4$ occurs because the luminosity distance-redshift relation is steeper than $z^2$ at $z \gtrsim 1.4$, while the differential increase in volume probed, $dV$, rises less quickly with $z$ than in Euclidean space.)

The differences between the flux density distributions for different spectral indices, $\alpha$, are entirely attributable to differences in $dz/d S_\nu$.  Figure \ref{fig:StandardCandle} demonstrates how the spectrum modifies the differential event rate at low $S_\nu$.  For decreasing spectra, $\alpha >0$, the source is less luminous in its emission frame compared to a flat-spectrum source detected at the same frequency, so that at high redshifts progressively fewer events are detected relative to events with flat or rising spectra. 
The steep falloff evident in the $\alpha=-1$ curve at $S_\nu \lesssim 0.3$ in Figure \ref{fig:StandardCandle} is due to the high redshift of the events detected at this flux density: at $z \gg 5$ the increasing slope of the luminosity distance-redshift relation and the diminishing rate of increase of survey volume with $z$ eventually overwhelm the effect of the $k$-correction.

\subsubsection{Fluence counts for standard candles and standard batteries}

\begin{figure}
\centerline{\epsfig{file=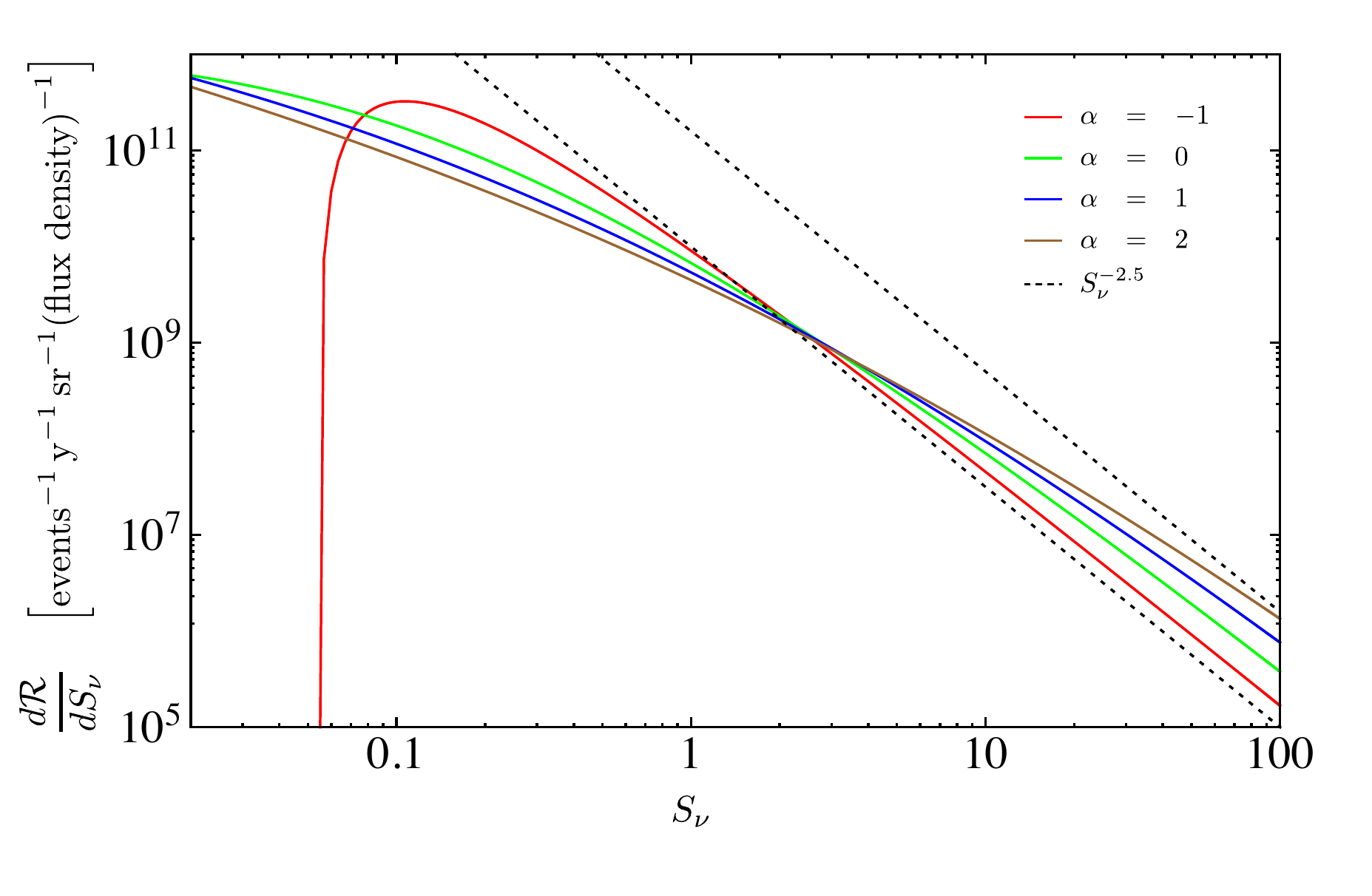,scale=0.4} \epsfig{file=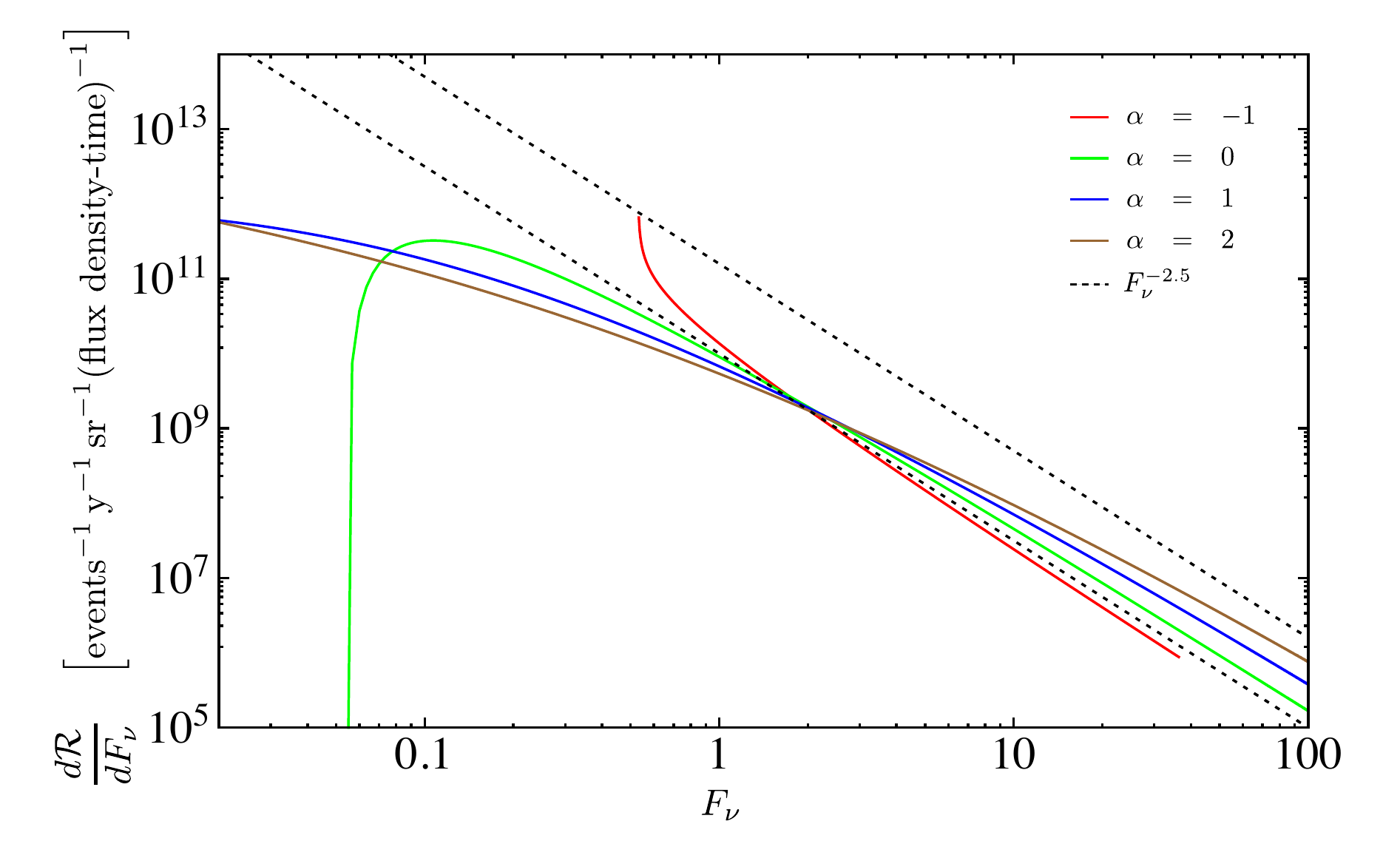,scale=0.42}}
\caption{Left: The differential event rate flux density distribution for a set of standard candles homogeneously distributed in space with no redshift evolution whose luminosity $L_0$ for each spectral index, $\alpha$, is chosen such that a $S_\nu = 1.0\,$ event occurs at $z=1$.  The normalisation of the overall event rate is set to one event per cubic Mpc per year ($R_0= 3.4 \times 10^{-68}\,$m$^{-3}$\,y$^{-1}$) so that the units of the differential counts are events\,y$^{-1}$\,sr$^{-1}$\,(flux density)$^{-1}$  .  
Right: the corresponding event rate fluence distribution for the same population, except that bursts are assumed to be standard batteries. The units of fluence are normalised so that an event at $z=1$ has a fluence of 1.0\,Jy\,ms (individually for each value of $\alpha$). The dotted lines in each plot are shown to illustrate the difference in slope between the derived distributions and a Euclidean distribution index of $-2.5$.  The two lines act cover the entire range of $\alpha$ shown.
} \label{fig:StandardCandle}
\end{figure}

\begin{figure}
\centerline{\epsfig{file=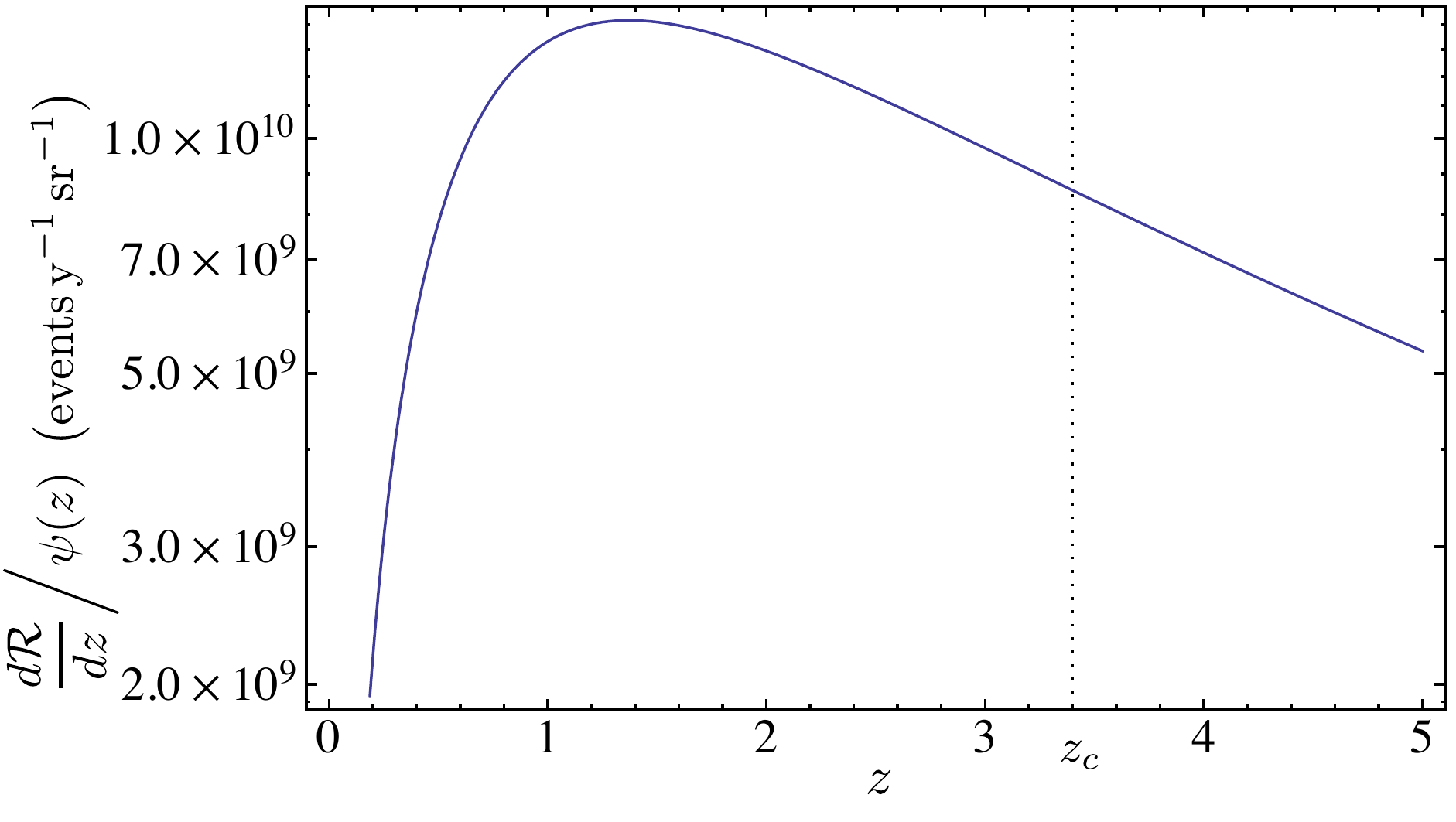,scale=0.4}}
\caption{The differential event rate redshift distribution for a set of standard candles for a survey whose limiting flux density is $S_0$. Here the normalisation of the overall event rate is set to the same value as in Figure \ref{fig:StandardCandle}: one event per cubic Mpc per year ($R_0= 3.4 \times 10^{-68}\,$m$^{-3}$\,y$^{-1}$).  The distribution cuts off at a redshift $z_c$, given by the solution to eq.(\ref{zcDefn}), that relates directly to the minimum detectable flux density.} \label{fig:StandardCandleZ}
\end{figure}

The corresponding fluence distribution for a population of standard candles is derived from the foregoing results by reinterpreting the ordinate of Figure \ref{fig:StandardCandle} as $F_\nu/\Delta t_e$ and making the substitution $\alpha \rightarrow \alpha - 1$ (see \S\ref{sec:countsgeneral}).  For instance, the fluence distribution for bursts which follow a flat spectrum, (i.e. $F_\nu \propto \nu^0$), is given by the $\alpha=1$ curve in the left panel of Figure \ref{fig:StandardCandle}.  However, one need not assume that the events occur with identical durations.  

We now introduce a new terminology, ``standard batteries'', since it is physically important to distinguish between a power and an energy.  There is an additional factor of $t$ difference and this is important because the cosmological time dilation changes these relations at high redshift.   If the FRBs radiate as ``standard batteries'' in which the total energy -- instead of the luminosity -- per unit bandwidth is assumed constant, the energy function is
\begin{eqnarray}
\Theta_E (E_\nu;z) = R_0 \psi(z) \delta (E_\nu - E_0). \label{eq:battery}
\end{eqnarray}
The direct relationship between the flux density and fluence statistics (viz. eqns.(\ref{SourceRate}) \& (\ref{SourceRateF2})) enables us to derive the fluence statistics for standard batteries simply by making the replacements $L_0 \rightarrow E_0$, $S_\nu \rightarrow F_\nu$ and $\alpha \rightarrow \alpha - 1$ in eqns.\,(\ref{StandardC})-(\ref{zcDefn}). For the sake of completeness, we demonstrate this correspondence explicitly by presenting the fluence counts computed from the standard battery assumption of eq.(\ref{eq:battery}) in the right panel of Figure \ref{fig:StandardCandle}.

%

\subsection{Broad power-law luminosity function}

Here we consider the event rate statistics for a population of events whose luminosity distribution follows a power law.  As just discussed, the use of the simplifying standard candle (or standard battery) model is a useful step to elucidate the geometrical effects of volume and distance on the source counts, but in reality the luminosity function is expected to be very broad.  If we interpret the bulk of the DMs observed in FRBs as a measure of distance we already know that the range in luminosity for FRBs varies by a factor of nearly $10^4$ as we go from the repeating source, FRB\,121102, which is the faintest FRB, to the Lorimer burst \citep[see, e.g. FRBCAT, ][]{Petroffetal2016}.  The range of distances for all known FRBs is only a little more than a factor of 10 so the distribution of the flux density or fluence, and hence the overall behaviour of the event rate counts will be dominated by the shape of the luminosity function and not by the distribution of FRBs in the Universe.

Once the importance of the luminosity function and its potential evolution with distance is recognised, the event rate counts become a powerful tool to constrain possible models of the population being counted.  Analysis of the count rate has the advantage that a well-defined complete sample can be used without precise knowledge of the distances for each individual object.  For FRBs this analysis is not affected by the distance uncertainty caused by the unknown contribution of the host DM or by clumpiness in the intergalactic baryon distribution \citep{McQuinn2014}. 

The general behaviour of the rate distribution is well known from studies of other astrophysical populations. \citet{vonHoerner1973} presents a full treatment of the behaviour of the source counts in the context of AGNs, and explains its evolution as a function of flux density (or fluence). Here we summarise those results.  It is useful to consider the differential rates in terms of a distribution normalised by the Euclidean counts slope of $S_\nu^{-5/2}$.  Following von Hoerner's description, we can divide the normalised counts into three regions rather than trying to describe it as a single power law:
\begin{itemize}
\item a maximum with a broad flat top, 
\item a section rising to the maximum as we decrease from the strongest sources, and 
\item a gradual decline from the maximum as we go to yet fainter flux densities.  
\end{itemize}
The portion at high flux densities requires evolution independent of any realistic cosmological model \citep[see][]{Wall83}, while the width of the maximum and the decrease to lower flux density depends more on the luminosity function than the cosmological distribution and evolution of the population.  Note that for surveys of non-transient phenomena such as the AGN, the statistics can only be improved by surveying to lower flux densities to increase the number of sources, since the whole sky is already measured at high flux densities.  However, for transient phenomena such as FRBs the statistics at high flux densities will always improve with time, so it is very useful strategy for future observations to target this area of parameter space requiring a large field of view, rather than higher sensitivity. 

The \cite{vonHoerner1973} analysis, using a range of luminosity functions, indicates that it is the bright end of the counts that provides the strongest constraints on the spatial distribution of the population.  For example, it is impossible to obtain a differential slope steeper than $-2.5$ (or integral slope steeper than $-1.5$) without either including the evolution of the density or luminosity to higher values in the past, unless one invokes strong local inhomogeneity (e.g. with the observer sitting in a local hole).  The counts then pass through an intermediate region where the slope decreases either slowly (making a very broad maximum in the Euclidian normalised counts) if the luminosity function is broad, or more sharply if it is narrow.  This regime is most sensitive to the width of the luminosity function, depending on how close the luminosity function is to a critical slope \citep[as discussed in Section IIb of][]{vonHoerner1973}.  The `critical value' of a power law luminosity function occurs when the increasing numbers of fainter sources exactly cancel the decreasing volume in which they can be seen.  In this situation, which is nearly the case for radio galaxies and quasars, there is no relation between flux density and distance (this is sometimes referred to as having no Hubble relation) and near the critical value the flux density - distance relation will have a very large scatter.  For luminosity fluctuations either side of this critical value there is either a direct or even an inverse flux density - distance relation in which fainter objects are more likely to be closer!  A rather extreme version of this occurs for extragalactic radio source surveys where the brightest known sources are AGN at large distances while the faintest radio sources are much more likely to be nearby starburst galaxies.


A complete and exact treatment of the event count statistics for a cosmological population is not fully analytically tractable.  The requirement to explicitly numerically integrate the event rate equation in order to determine the behaviour of the event rate statistics obliges us to specify the redshift evolution of the population.  In particular, in the present treatment we discuss a set of simple models in an attempt to encompass a range of possible representations of the FRB population redshift evolution.  In particular, we employ the source counts formalism to discuss in detail a specific class of models that may apply to FRBs, in which the progenitor population is coupled to some power of the star formation rate. Informed by these results, we then develop further a physical understanding of the behaviour of the event count statistics.    

The motivation for a studying this more restricted range of progenitor scenarios in detail here is to investigate whether any such model is compatible with a steep rate counts distribution that may be suggested by present data (Paper I) and, if so, to deduce what this in turn would imply about the FRB population.

\subsubsection{Physical models for abundance evolution}

If the population of FRBs extends to greater distances, as we have assumed, than any realistic model of FRB statistics must incorporate potential evolution in the event rate density throughout cosmic history.
We incorporate evolution by specifying a realistic prescription for the redshift dependence of $\Theta_L(L_\nu;z)$ and $\Theta_E(E_\nu;z)$.  We restrict the discussion to only the set of models in which the rate density of bursts varies with redshift, and there is no evolution of the shape of the luminosity distribution with redshift.  Though not rigorously justifiable in general, this is plausible for FRBs on the basis that the luminosity function for some compact events is expected to depend on local redshift-independent physics (i.e. the electrodynamics and quantum mechanics of the emission mechanism), whereas the overall event rate is set by the evolution of the progenitor population.  Thus the rate density luminosity function is assumed here a separable function of $L_\nu$ and $z$,
\begin{eqnarray}
\Theta_L(L_\nu;z) = \theta_L(L_\nu) \psi(z) \hbox{ and } \Theta_E(E_\nu;z) = \theta_E(E_\nu) \psi(z). \label{FullPowerLaw}
\end{eqnarray}

A simple generic model for the redshift evolution of a population is characterised by at least four parameters: the (normalising) rate per comoving volume at the present epoch, $R_0$, the slope of the increase in rate density, $dR/dz \equiv R_{+}'$, up to some peak redshift $z_{\rm p}$, and the slope, $dR/dz \equiv R_{-}'$ of the decline in abundance at $z>z_{\rm peak}$.  
At present, the paucity of existing FRB event count data pushes the exploration of such a four-dimensional model well beyond the scope of this work.  

We instead restrict the discussion here to a relatively simple formulation that is still of physical relevance to the problem, namely a rate based on a progenitor population whose abundance is governed by the evolution of stellar processes throughout the history of the Universe.   Specifically, we consider a two parameter formulation in which the rate is related by some constant of proportionality, $K$, to some power, $n$, of the star formation rate per comoving volume (SFR),
\begin{eqnarray}
\psi(z) = K \Psi(z)^n.
\end{eqnarray}

Of special significance are models in which $n=1$, corresponding to an evolutionary history that is linearly proportional to the SFR, and one in which $n=2$ which would, for instance, correspond to events whose progenitors might involve two stars that formed independently (e.g. some recycled neutron star systems in a dense cluster environment, or the mergers of two stars that formed independently).
We adopt throughout this paper the following prescription for the cosmic star formation rate (Madau \& Dickinson 2014):
\begin{eqnarray}
\Psi(z) = 0.015 \frac{(1+z)^{2.7}}{1 + [(1+z)/2.9]^{5.6}} {\rm M_\odot}\,{\rm y}^{-1}\,{\rm Mpc}^{-3}. \label{MadauSFR}
\end{eqnarray}
In the prescription given by eq.(\ref{FullPowerLaw}) the event rate is then fully specified, and the constant $K$ represents the conversion efficiency between the mass rate of star formation and the FRB rate.  The units of $\Theta(L_\nu;z)$ are events per volume per luminosity per time, so for $n=1$, the units of $K$ are events per unit solar mass of star formation (i.e. events\,${\rm M}_\odot^{-1}$).
 More generally, for $n \neq 1$, the constant $K$ has units of events\,${\rm M}_\odot^{-n}$ y$^{n-1}$\,Mpc$^{3n-3}$.  
 
It is instructive to to provide an example of a physical interpretation for the value of $K$ for $n=2$.  Suppose that the FRB rate depends on the interaction between two stars that formed independently.  Then the FRB rate per unit volume depends on the product of the formation rate of stars of type 1 with the formation rate of stars of type 2 and with the interaction cross-section, $\sigma_{12}$, between the two types: 
\begin{eqnarray}
\hbox{FRB rate density}\, [\hbox{events}\,{\rm y}^{-1}\,{\rm Mpc}^{-3}] = \sigma_{12} [A_1 \psi(z)] [ A_2 \psi(z) ],
\end{eqnarray} 
where $A_1$ (with units stars$\,{\rm M}_\odot^{-1}$) is a conversion efficiency that specifies how many stars of type 1 form per unit mass of star formation and similarly $A_2$ is the corresponding efficiency for stars of type 2.  The interaction cross-section, $\sigma_{12}$, has units of (events/stars$^2$)\,y$^1$\,Mpc$^3$.  The constant $K$ is therefore the product of the star formation efficiencies and a cross section: $K = A_1 A_2 \sigma_{12}$ and hence has units of M$_\odot^{-2}$\,y\,Mpc$^{3}$.


One can envisage a number of modifications to the evolutionary model presented here. 
An obvious modification involves accounting for the fact that the events responsible for FRBs may represent the endpoint of stellar evolution, and they thus require a finite time to evolve from the epoch at which their progenitor stars form to their manifestation as FRBs.  Thus $\psi(z)$ may in turn be offset in time from the SFR.  We do not explicitly take this delay into account in the present modelling because the scant information on FRB counts is not yet of sufficient quality to merit the inclusion of this additional parameter, but we note that it appears to be important in modelling the progenitors of some other explosions, where a distribution of formation-to-burst times is adopted (e.g. GRBs; see Shin'ichiro 2004).  

We remark that the $n=2$ scenario is remarkably similar in form to the evolution function used to fit models of the evolution of powerful AGN, for which the density is often modelled to scale as $(1+z)^6$ up to a cutoff redshift of $z \approx 2.5$ \citep[e.g.][]{Schmidt72}.  Thus this particular scenario possesses the virtue of elucidating the source counts expected if the evolution of the FRB population is linked to the AGN/supermassive black hole phenomenon. Moreover, since we do not yet have good constraints on the evolution of the FRB population, an important additional motivation for investigating the $n=2$ scenario is that it is a convenient means to investigate populations whose evolution occurs more rapidly than the SFR. Exploration of this model against the $n=1$ case shows how the rate of evolution is reflected in the source counts behaviour.
 
Finally, we add that the $n=1$ case was discussed by \cite{CordesWasserman2016} in the context of neutron star-based FRB models. The two formalisms conceptually agree (e.g. in redshift-dependent terms), and identify similar qualitative behaviour in the cases that overlap \citep[as summarised in the three points below eq.(68) in][]{CordesWasserman2016}.  Here we broaden the scope of the flux-density counts investigated by \cite{CordesWasserman2016} to include the fluence counts, which notably incorporates a difference in the $k$-correction. A broader range of luminosity distributions and spectral indices are investigated here.  

\subsubsection{Power-law luminosity functions} \label{sec:powerlawtreatment}

We now turn our attention to the case in which the events span either a range of luminosities or energies.  We take the distributions to possess cutoffs at luminosities $L_{\rm min}$ and $L_{\rm max}$, or energies $E_{\rm min}$ and $E_{\rm max}$, respectively.  For a given value of $S_\nu$ or $E_\nu$ events are detectable over the range of redshifts, $[z_{\rm min},z_{\rm max}]$ set by the limits on the luminosity or energy ranges.  We investigate the cases in which the luminosity and energy functions follow a power law with index $\gamma$:
\begin{eqnarray}
\theta_L(L_\nu) = \frac{\gamma-1}{L_{\rm min}^{1-\gamma} - L_{\rm max}^{1-\gamma}} L_\nu^{-\gamma} \hbox{ and }
\theta_E(E_\nu) = \frac{\gamma-1}{E_{\rm min}^{1-\gamma} - E_{\rm max}^{1-\gamma}} E_\nu^{-\gamma}.
\end{eqnarray}
We currently have poor constraints on the values of $L_{\rm max}$ and $E_{\rm max}$ for the FRBs but these upper limits may be regarded as parameters necessary to define the limits of some of the integrations encountered below.

The resulting flux density and fluence distributions are written assuming power-law distributions in $L_\nu$ and $E_\nu$ respectively to obtain
\begin{subequations}
\begin{eqnarray}
\frac{d {\cal R}}{dS_\nu} &=& {\cal Y} (L_{\rm min},L_{\rm max},\alpha,\gamma), \label{SnuPowerLaw} \\
\frac{d {\cal R}}{dF_\nu} &=& {\cal Y} (E_{\rm min},E_{\rm max},\alpha-1,\gamma), \label{FnuPowerLaw} \\
\hbox{where  } {\cal Y} (X_{\rm min},X_{\rm max},\alpha,\gamma) &=& 4 \pi d\Omega \,  D_H^5 \frac{\gamma-1}{{X_{\rm min}^{1-\gamma} - X_{\rm max}^{1-\gamma}}} (4 \pi S_\nu)^{-\gamma} \int_{z_{\rm min}}^{z_{\rm max}} dz \left[ \frac{D_L(z)^2 }{(1+z)^{1-\alpha} } \right]^{-\gamma}  \psi(z)  \frac{(1+z)^{\alpha}}{E(z)} \left[ \int_0^{z} \frac{dz'}{E(z')} \right]^4.
\end{eqnarray}
\end{subequations}
Clearly, the differential flux density and fluence distributions are similar in form, with latter derived from the former using the replacements $S_\nu \rightarrow E_\nu$, $L_{\rm min,max} \rightarrow E_{\rm min,max}$ and $\alpha \rightarrow \alpha-1$.  These relationships allow us to derive the fluence counts of eq.(\ref{FnuPowerLaw}) for a power-law energy function directly from the flux density counts for a power-law luminosity function. However, we stress that these two expressions do not represent equivalent physics: the assumption that the luminosity follows a power law distribution is in general not equivalent to the assumption that the energy follows a power law. Furthermore, the limits $z_{\rm min}$ and $z_{\rm max}$ are not identical.  For eq.(\ref{SnuPowerLaw}) and eq.(\ref{FnuPowerLaw}) the minimum/maximum redshifts are obtained by the solutions to, respectively,
\begin{eqnarray}
L_{\rm min/max} &=& \frac{4 \pi D_L(z_{\rm min/max})^2}{(1+z_{\rm min/max})^{1-\alpha}} S_\nu  \\
 E_{\rm min/max} &=& \frac{4 \pi D_L(z_{\rm min/max})^2}{(1+z_{\rm min/max})^{2-\alpha}} F_\nu 
\end{eqnarray}

\begin{figure}
\centerline{\epsfig{file=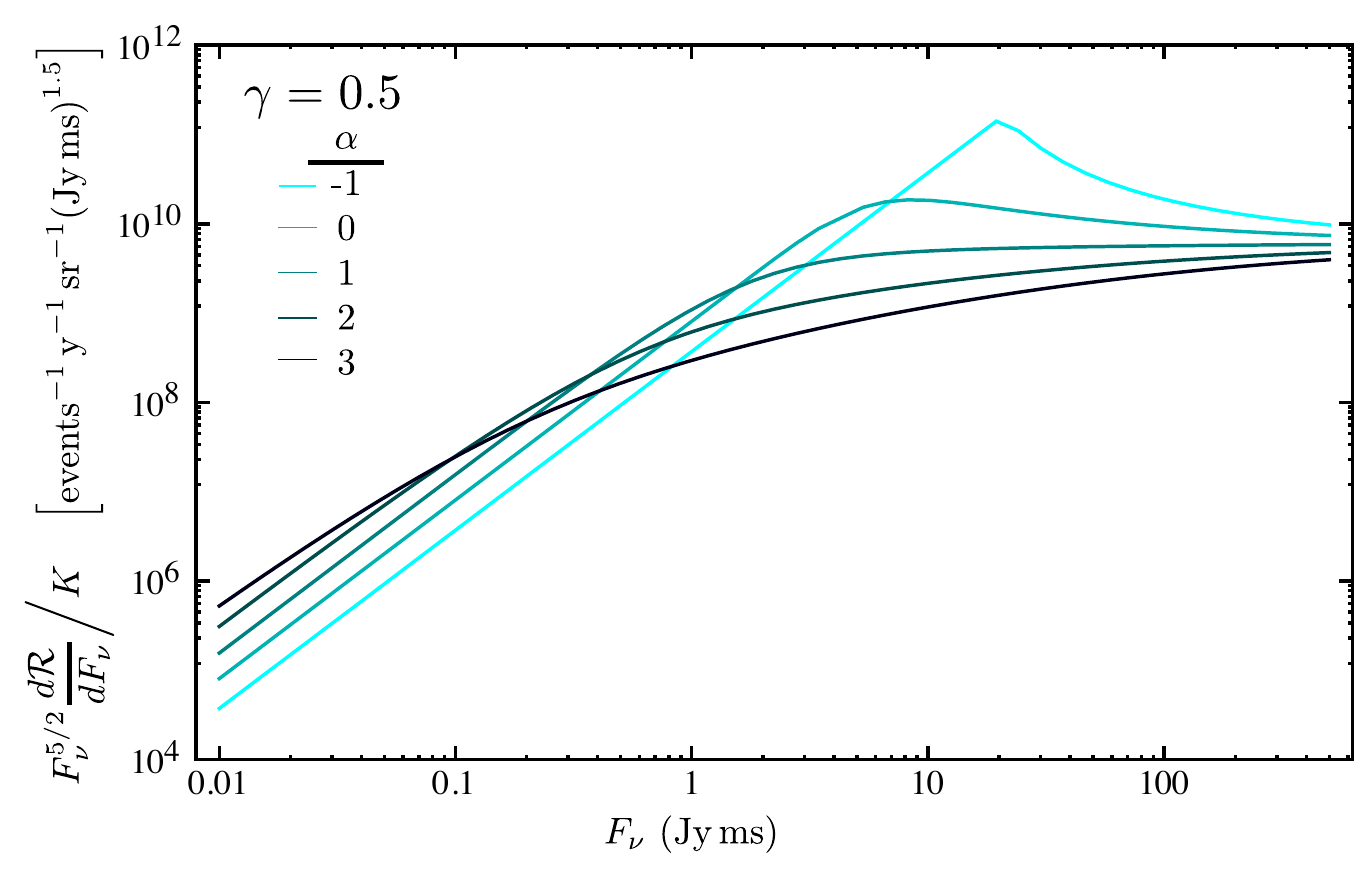,scale=0.6} \epsfig{file=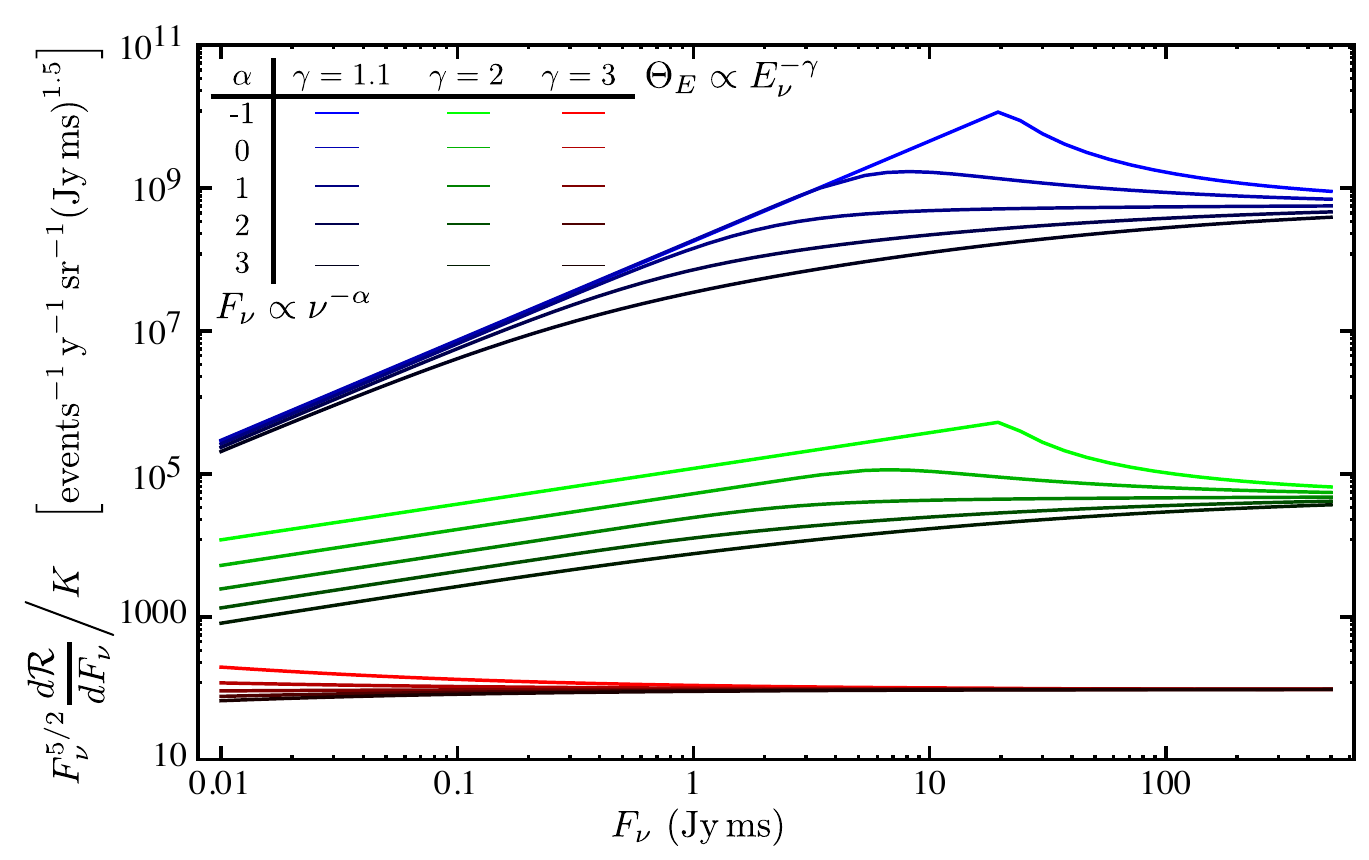,scale=0.6}}
\caption{Differential source counts for a power-law energy function with a variety of indices, $\gamma$, and for various burst spectral indices, $\alpha$, assuming that the abundance is linearly proportional to the SFR, with $\psi(z) = K \Psi(z)$.  The event rate curves are normalised by $F_{\nu}^{-5/2}$, the functional dependence on $F_\nu$ for a Euclidean Universe.  The overall rate is further normalised by the constant, $K$ events\,M$_\odot^{-1}$.   For both panels the limits of $\theta_E$, the energy function, were chosen such that a burst at $z=1$ with $\alpha=1$ ranges between $0.01\,$mJy\,ms and $10\,$Jy\,ms. The equivalent flux density distribution is derived reinterpreting the plot with $1$ subtracted from the value of $\alpha$, by re-interpreting the $x$-axis to be in Jy, and with the luminosity function chosen so that a burst at $z=1$ with $\alpha=0$ ranges between $0.01\,$mJy and $10\,$Jy.  A similar reinterpretation can be applied to Figures \ref{fig:LuminositySFRpartb} and \ref{fig:LuminositySFRsquared} to derive their corresponding flux density distributions.
} \label{fig:LuminositySFR}
\end{figure}

\begin{figure}
\centerline{\epsfig{file=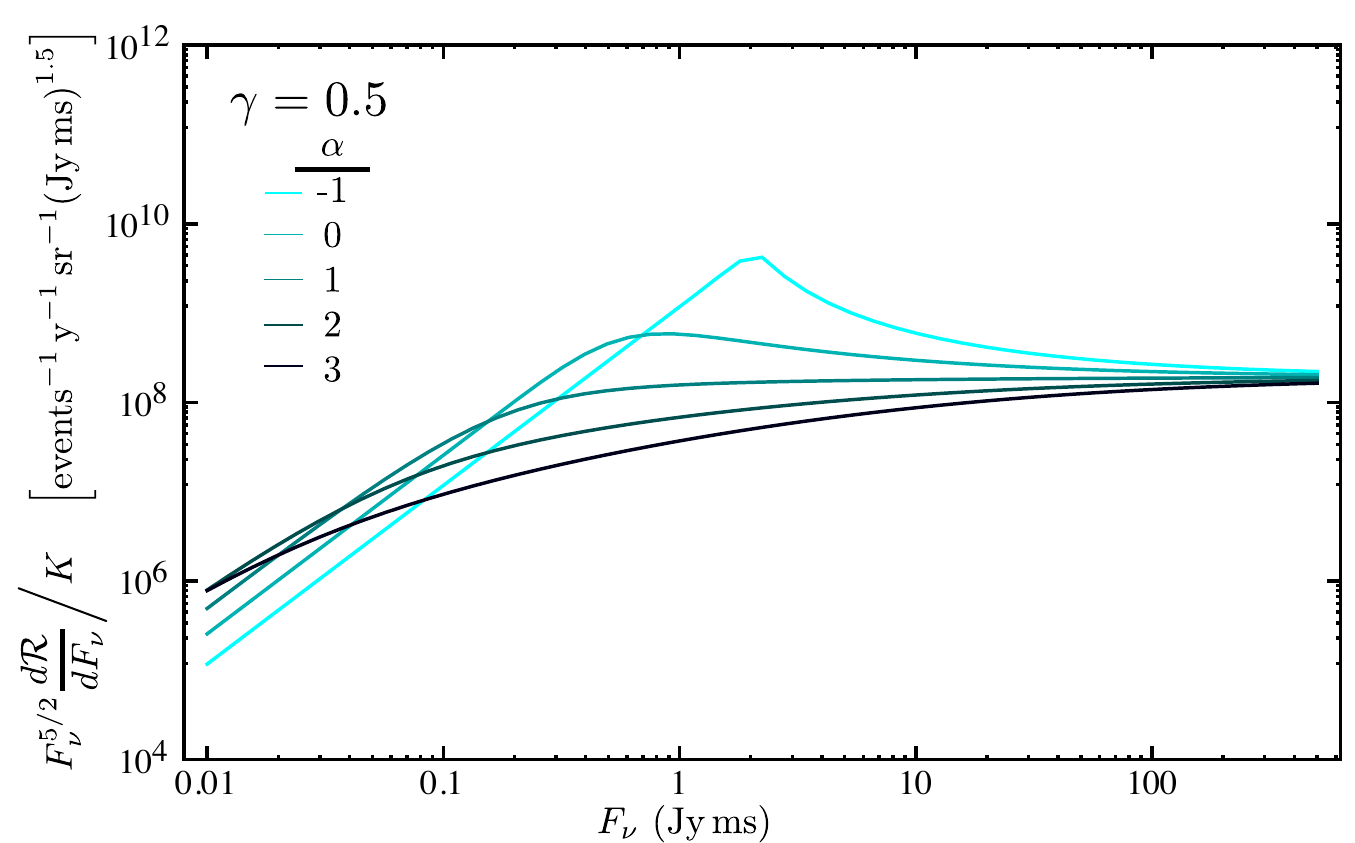,scale=0.6} \epsfig{file=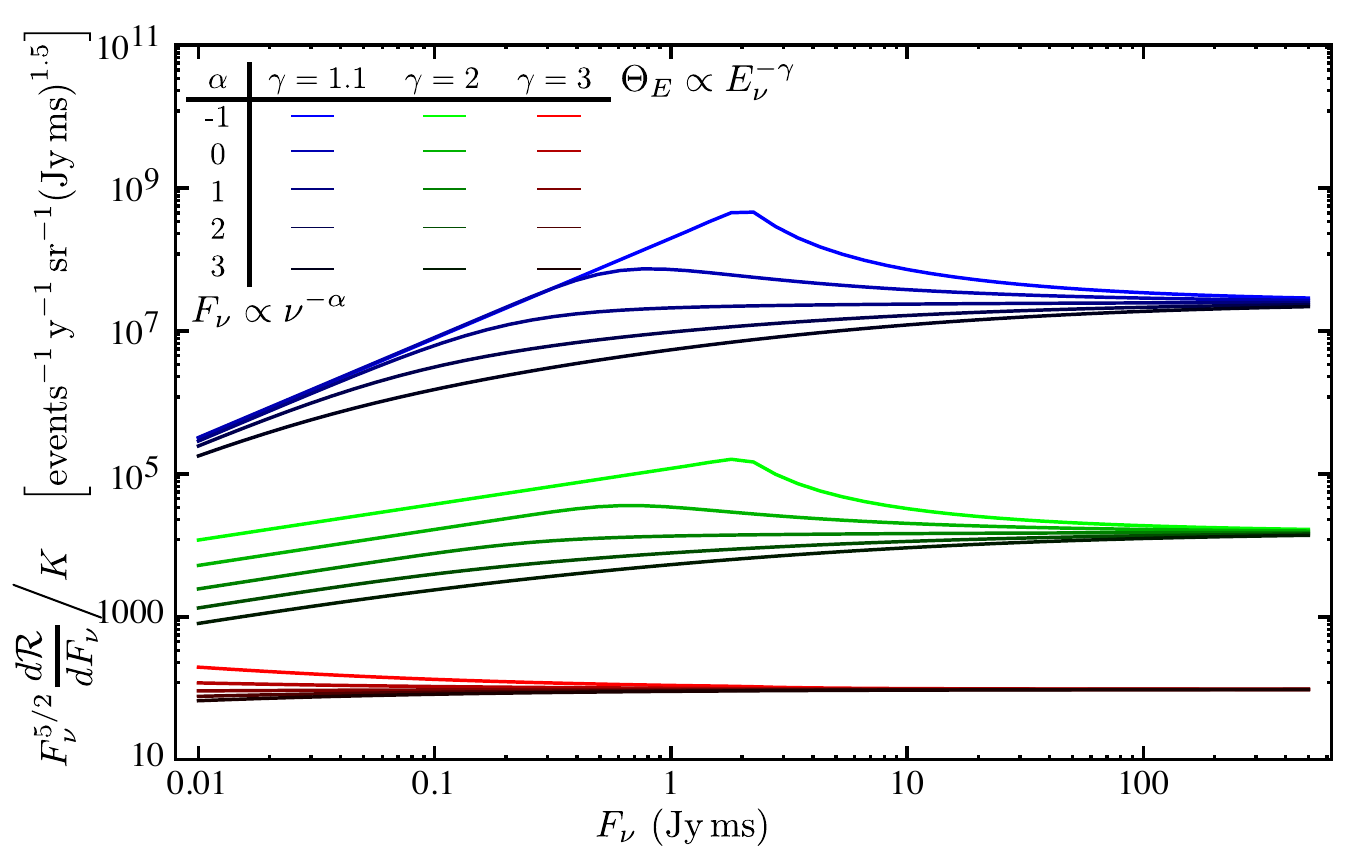,scale=0.6}}
\caption{The same as Figure \ref{fig:LuminositySFR} except that $E_{\rm max}$ has been shifted to a value a factor of 10 lower.  This results in a corresponding factor of ten shift in the point in $F_\nu$ at which the curves break, demonstrating that the break is due to the value of $E_{\rm max}$. } \label{fig:LuminositySFRpartb}
\end{figure}

\begin{figure}
\centerline{\epsfig{file=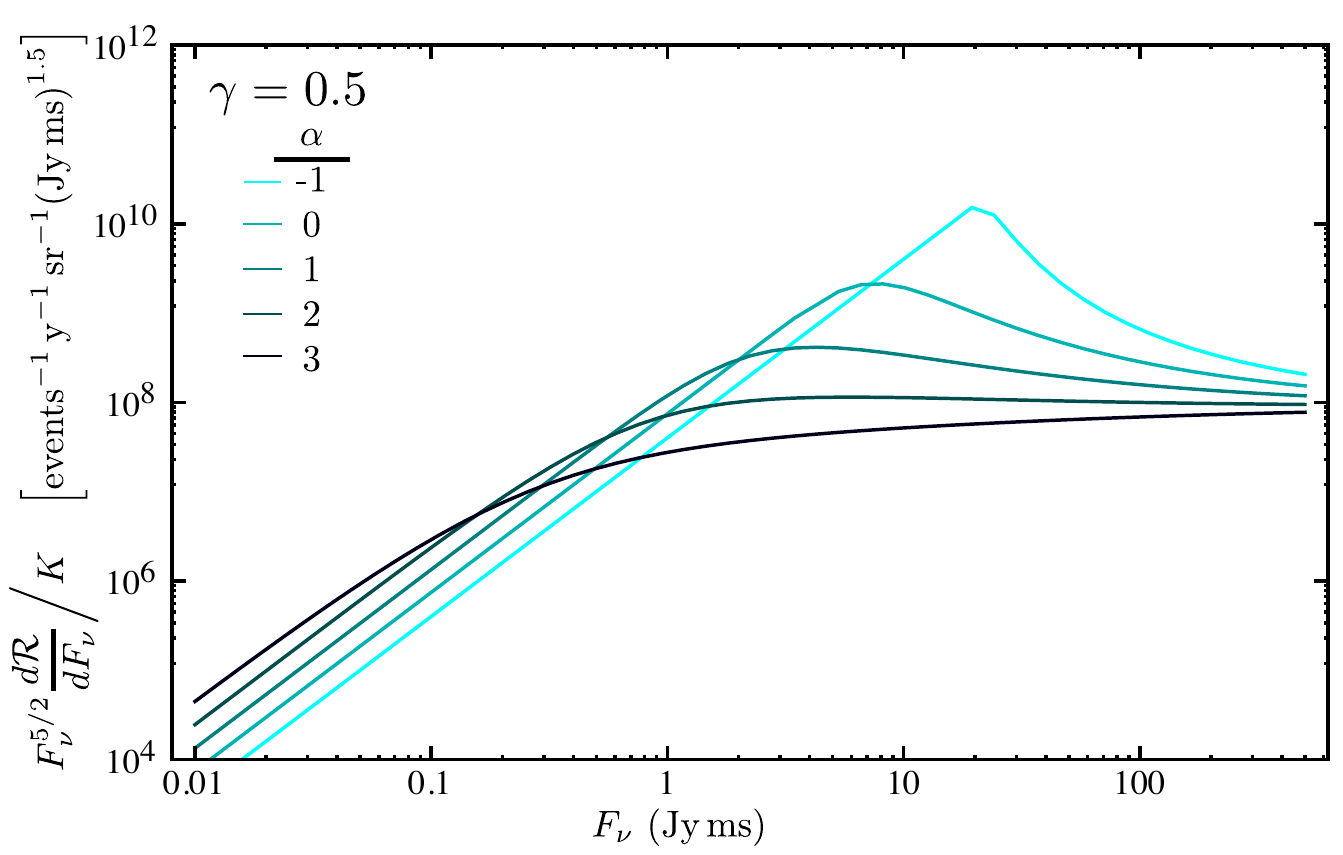,scale=0.6} \epsfig{file=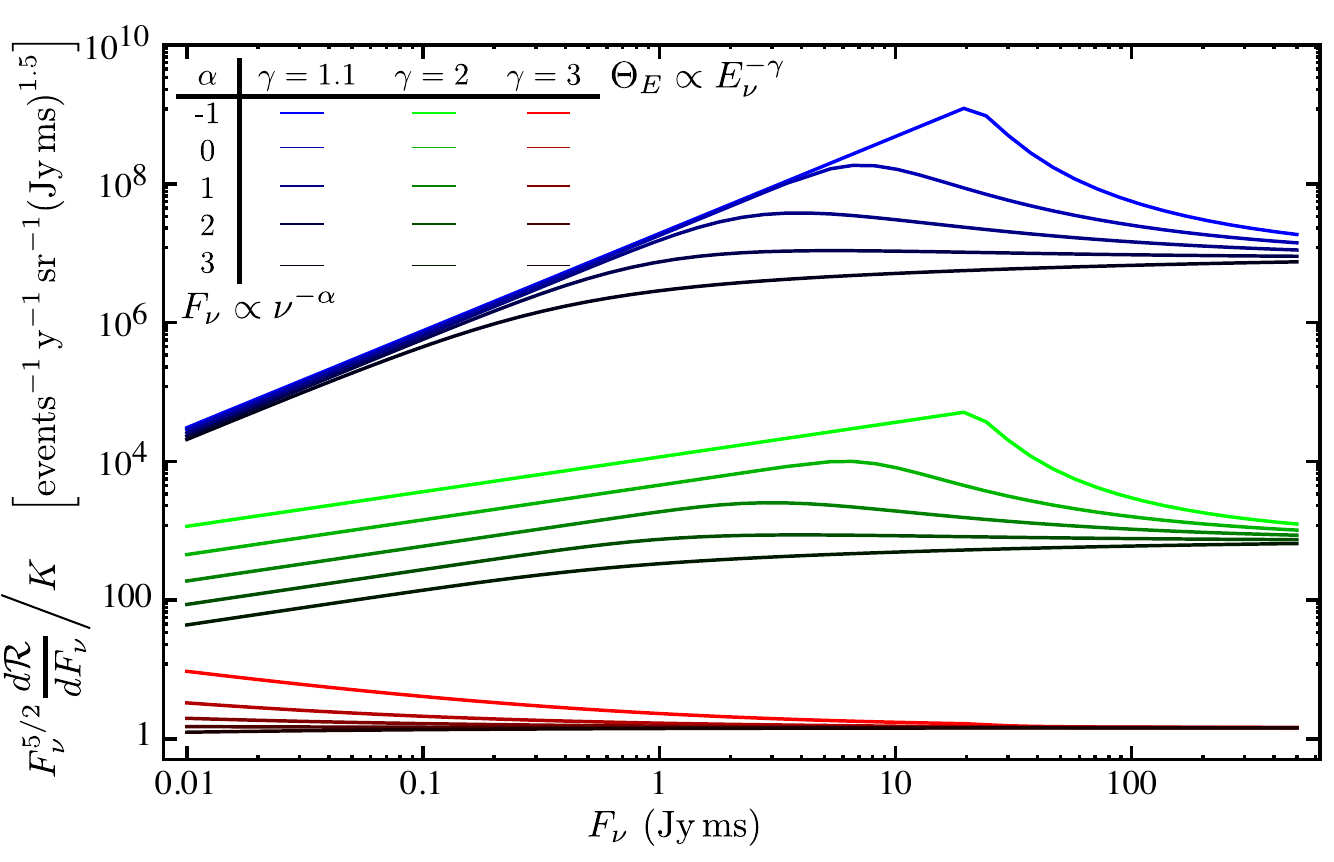,scale=0.6}}
\caption{The same as Figure \ref{fig:LuminositySFR} except that the relative abundance of the events as a function of redshift is taken to scale as the square of the SFR, $\psi(z) = K \Psi(z)^2$, with the conversion efficiency $K$, expressed in events\,M$_\odot^{-2}$y\,Mpc$^3$, normalised out of the rate. For luminosity functions substantially flatter than $E_\nu^{-2}$ (e.g. left panel) the distribution exhibits a high-$F_\nu$ break associated with the maximum energy, $E_{\rm max}$.  The flatter the distribution, the steeper the source counts at flux densities above the break.  The left plot demonstrates the steepening associated with an extremely flat ($\gamma=0.5$) energy function. For steeper distribution functions, $\gamma=3$, there is little difference in the curves for any of the values of $\alpha$ plotted.
} \label{fig:LuminositySFRsquared}
\end{figure}

These equations are integrated for each specific value of $S_\nu$ (or $F_\nu$) over the range of interest for given values of $\alpha$, $\gamma$, $L_{\rm min}$ and $L_{\rm max}$.  The results are shown in Figs.\,\ref{fig:LuminositySFR}-\ref{fig:LuminositySFRsquared}.  A summary of these results is as follows:


\begin{itemize}
\item[1.] There is a break in the source count distribution at $F_\nu \approx E_{\rm max} (1+z_0)^{2-\alpha}/(4 \pi D_L(z_0)^2)$ (or $S_\nu \approx  L_{\rm max} (1+z_0)^{1-\alpha}/(4 \pi D_L(z_0)^2)$ in the flux density distribution), where $z_0$ is the characteristic redshift at which there is a turn-over in the distribution of the population.  The origin of the break can be understood in terms of the sharp decline in abundance of sources at redshifts below the peak of the SFR.  If the event abundance were to cut off at a redshifts $z>z_0$, then the source count distribution would cut off at $F_\nu = L_{\rm max} (1+z_0)^{2-\alpha}/(4 \pi D_L(z_0)^2)$.  To the extent that the abundance declines smoothly from its maximum to $z=0$, the decline in source counts to higher flux densities is less abrupt.

This interpretation is borne out in Figure \ref{fig:LuminositySFR}.  For this plot the value of $E_{\rm max}$ ($L_{\rm max}$) was set to be equivalent to a 10\,Jy\,ms (10\,Jy) source at a redshift of $1$.  Since the abundance (i.e.\,the SFR) peaks at $z_0 \approx 1.5$, the break is predicted at a flux density of $\approx 3\,$Jy.  This predicted flux density corresponds to the observed flux density of the break in the source count distribution. Figure \ref{fig:LuminositySFRpartb} further demonstrates clearly that the break is directly associated with $E_{\rm max}$ by showing that a factor of 10 decrease in $E_{\rm max}$ results in a commensurate factor of 10 decrease in the break flux density.  

Of course, we do not know the value of $E_{\rm max}$ or $L_{\rm max}$ {\it per se}, but as observations improve we would expect to see a break in the counts, and the foregoing analysis provides a means of interpreting its location.

\item[2.] For shallow luminosity functions, $\gamma<2$, the source count distribution closely matches the slope of the luminosity function at low flux densities but with some dependence on the $k$-correction.  This can be understood as an effect of the sharp peak in abundance of events with redshift.  In the limit in which all the events are confined to an extremely narrow redshift range (i.e.\,they all occur at the same distance), then the slope of the source count distribution will be an exact match to the slope of the luminosity function. For distributions steeper than $\gamma=2$, they match the Euclidean source counts index at low flux densities.

\item[3.] For $\gamma<2$, the source counts are steeper than Euclidean for at least a decade in $F_\nu$ above the break and for these distributions
the post-break slope is steeper for more sharply-peaked abundance functions.  This is demonstrated in Figures \ref{fig:LuminositySFR} and \ref{fig:LuminositySFRsquared}, which shows that the post-break slope of the source counts is steeper when the abundance is quadratically proportional to the SFR instead of linearly proportional.  For fluences above the break the source counts probe only those events whose redshifts are less than $z_0$.  Progressively higher fluences probe the event redshift distribution at progressively lower redshifts.  The abruptness of the decline in the source counts is directly related to the abruptness in the decline of the abundance of events as $z$ decreases towards zero.

\item[4.] The post-break counts decline more steeply the flatter the luminosity function. This is highlighted in the left panel of Figure \ref{fig:LuminositySFRsquared}, where we see that an exceptionally flat luminosity function gives rise to steep source counts.  However, for luminosity functions that decline more steeply than $\gamma=2$, we see that all source count distributions decline as $F_\nu^{-2.5}$ above the break, as if the source counts follow a Euclidean distribution.   Furthermore, we see that at sufficiently high $F_\nu$ all distributions asymptote towards the Euclidean scaling of $F_\nu^{-5/2}$.  This relates to the fact that flat luminosity distributions preferentially detect objects at high redshifts, whereas steep luminosity distributions primarily probe the component of the population at low-redshift ($z \ll 1$).
\end{itemize}

%

\subsection{Explanation of Source Counts Behaviour}
The objective in this section is to explore the behaviour of the source counts with particular reference to the underlying redshift distribution of the events detectable at any given fluence.  This is a prelude to the discussion of the redshift and DM distributions in the following section. The basis of our explanation is the differential redshift distribution re-written in the form (cf.\,eq.\,(\ref{SourceRateF2})):
\begin{eqnarray}
\frac{d^2{\cal R}}{dF_\nu dz} = \frac{(4 \pi)^{1-\gamma} D_H (\gamma-1)}{E_{\rm min}^{1-\gamma} - E_{\rm max}^{1-\gamma}} d\Omega \, \left[ \left( \frac{D_L(z)^2 F_\nu}{(1+z)^{2-\alpha}} \right)^{-\gamma} \psi(z) \frac{(1+z)^{\alpha-1}}{E(z)} D_M(z)^4 \right], \,\,\,  \frac{(1+z)^{2-\alpha}}{4 \pi D_L^2}  E_{\rm min} < F_\nu < \frac{(1+z)^{2-\alpha}}{4 \pi D_L^2}  E_{\rm max}, \nonumber \\ \label{RedshiftDistn}
\end{eqnarray}
where the distribution is zero outside the region $z_{\rm min}(F_\nu) < z < z_{\rm max} (F_\nu)$, and all the redshift-dependent terms are grouped inside the square brackets. For readability, we shall explicitly only discuss the fluence distribution here, but all of the results are equally applicable to quantities involving the flux density, using the replacements $F_\nu \rightarrow S_\nu$, $\alpha \rightarrow \alpha +1$, $E_{\rm min} \rightarrow L_{\rm min}$ and $E_{\rm max} \rightarrow L_{\rm max}$.

We identify three regimes:  

\medskip

\noindent
i) $z_{\rm max} \ll 1$:  This is the regime which applies to source counts in the limit $F_\nu \rightarrow \infty$. The fluence associated with an object at redshift $z$ is $F_\nu \gg E_{\rm max} (1+z)^{2-\alpha}/(4 \pi D_L(z)^2)$, and the condition $z_{\rm max} \ll 1$ implies that $H_0^2 E_{\rm max}/ 4 \pi c^2 \ll F_\nu$.   Qualitatively we understand this as follows: as the sensitivity worsens, the detection threshold increases implying that larger amplitude bursts are detected.  $F_\nu \rightarrow \infty$, and the cutoff redshift out to which we observe events moves progressively closer to us, so that our source counts become sensitive to events only in the local Universe.  

In the $z \ll 1$ regime the approximation to the source counts given by eq.(\ref{lowzCounts}) therefore applies.  
Now, in this regime the star formation rate is approximated by $\Psi(z) \propto (1+z)^{2.7}$ which is constant in the limit $z \ll 1$ to lowest order in $z$.  This is to say that there is negligible evolution over the range of redshifts probed, so the luminosity function $\Phi(L_\nu,z)$ is a function of $L_\nu$ only.  This immediately implies that the source counts in the high flux density limit are Euclidean, as shown explicitly in \S 3.1.

\medskip \noindent
ii) $z_{\rm max} \gg 1$: This condition corresponds to the behaviour of source counts in the low fluence regime.  For $z\gg 1$, $D_L > c z/H_0$ which in turn implies the approximate relation $F_\nu \ll H_0^2 L_{\rm max}/ 4 \pi c^2$.  

\begin{figure}
\centerline{\epsfig{file=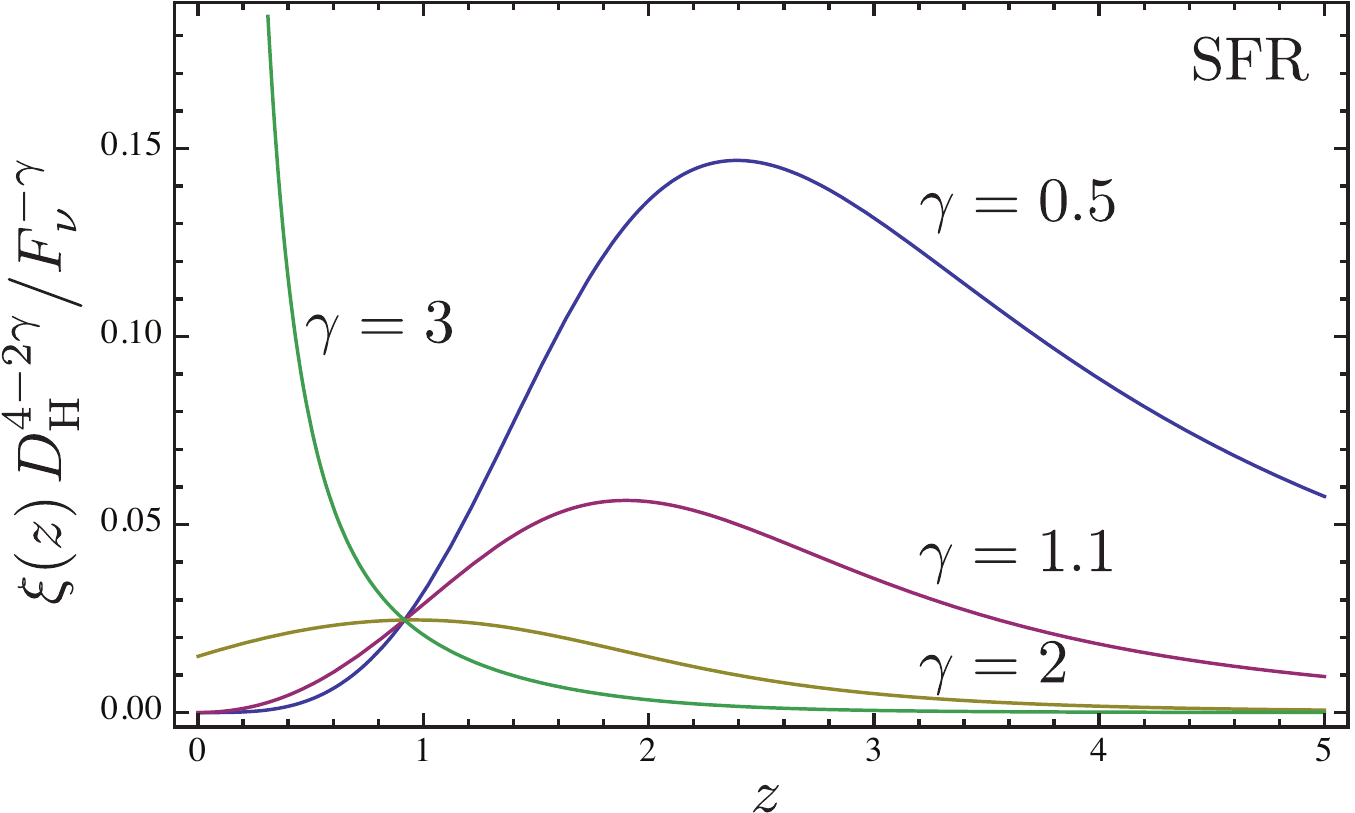,scale=0.6} \epsfig{file=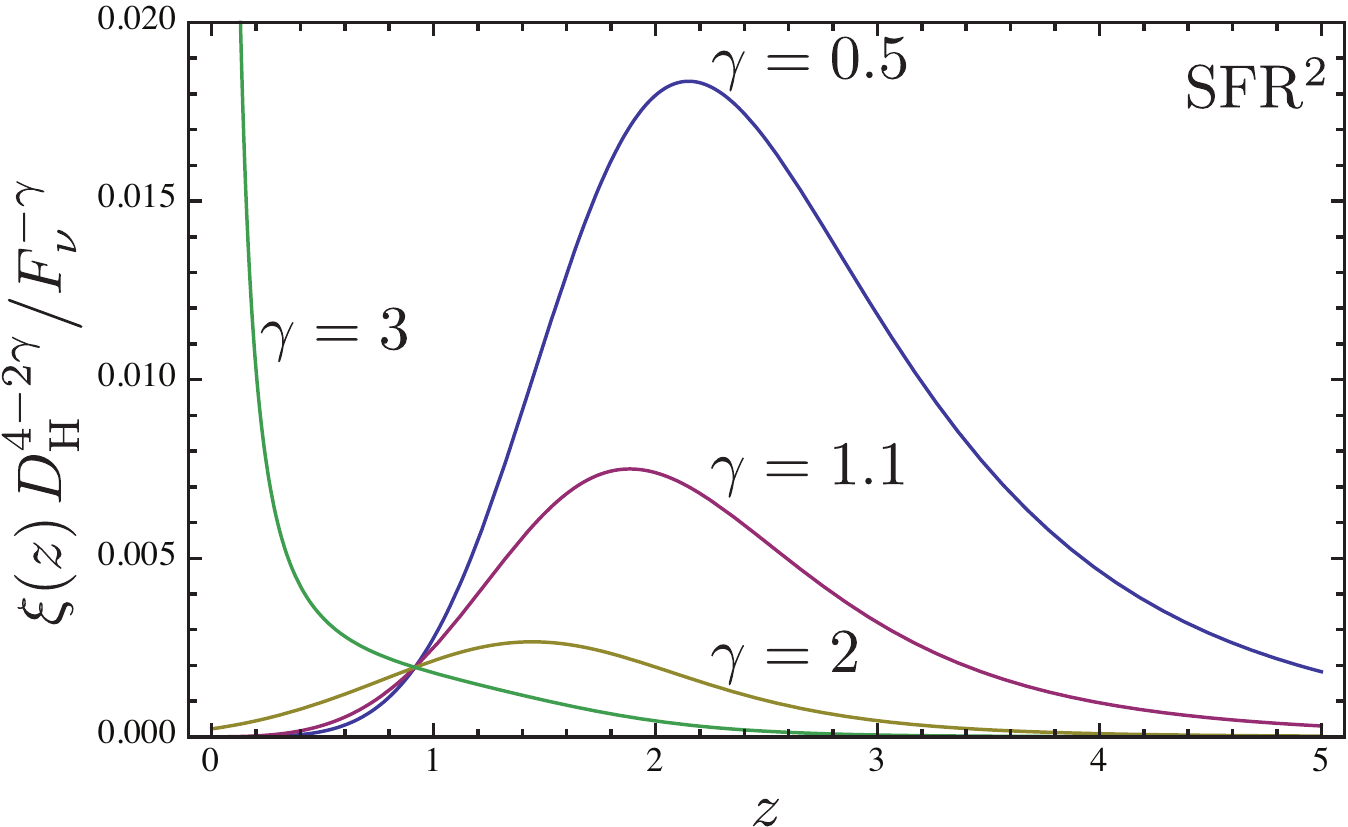,scale=0.6}}
\caption{The differential redshift and flux density distribution of eq.\,(\ref{xiDefn}) for a variety of $\gamma$ indices for $\alpha=1$ and for abundance evolution scenarios that scale either (left) linearly or (right) quadratically with the SFR.  The plot is normalised by the constant $F_\nu^{-\gamma} D_H^{4-2 \gamma}$ for plotting purposes.} \label{fig:xiplot}
\end{figure}

Under the assumption that $E_{\rm min} \ll E_{\rm max}$, the limits of the redshift distribution, $z_{\rm min}$ and $z_{\rm max}$, are therefore unimportant in this regime, and the distribution of the events that contribute to the source counts at $F_\nu \ll H_0^2 L_{\rm max}/ 4 \pi c^2$ is determined by the redshift-dependent terms in eq.(\ref{RedshiftDistn}), which are:
\begin{eqnarray}
\frac{d^2 {\cal R}}{dF_\nu dz} \propto F_\nu^{-\gamma} D_M(z)^{4-2 \gamma} \,  \psi(z)\,  \frac{(1+z)^{\alpha(1-\gamma)-1}}{E(z)} \equiv \xi(z). \label{xiDefn}
\end{eqnarray}
The behaviour of the function $\xi(z)$, as shown in Figure \ref{fig:xiplot}, is dominated by the term $D_M^{4-2\gamma}$.  The behaviour of the distribution changes at the point $\gamma=2$.  For $\gamma>2$ the redshift distributions are peaked at $z \sim 0$, whereas for $\gamma < 2$ we see that the redshift distribution peaks at increasingly high redshift, so the source counts are dominated by high-redshift events.  

This explains the Euclidean source counts observed in the low-$F_\nu$ regime for distributions with $\gamma>2$.  The distribution of events is dominated by events at low redshift.  Thus the form of the source counts reverts to the $z\ll 1$ form embodied in eq.(\ref{lowzCounts}), and this again gives rise to a Euclidean source count distribution.   

However, for $\gamma<2$ the redshift distribution is biased towards preferentially detecting events at the highest redshifts.  The greater the tendency for events to accumulate over a range of higher redshifts, the greater the tendency for the source counts index to match that of the underlying luminosity function.  If the objects were {\it all} at an identical redshift, the luminosity distance would be identical for all objects, and there would be a one-to-one relation between event luminosities and flux densities, so that the source counts scale as $F_\nu^{-\gamma}$.  

This effect is also influenced by the $k$-correction: for $\alpha=-1$ the flux density of events at high redshift partially negates the flux density falloff with distance, causing the redshift distribution to be more strongly peaked to high redshifts.  However, as $\alpha$ increases, the flux densities of events at large distances fall progressively faster than $\propto D_L(z)^{-2}$, thus moderating the redshift bias.  For values $\alpha \gtrsim 0$ the bias is such that the events detected are more evenly distributed in redshift space relative to events with $\alpha < 0$, and the index of the source counts lies somewhere between $-\gamma$ and the Euclidean value of $-2.5$.

Another way to see that the distribution scales roughly as $F_\nu^{-\gamma}$ is that in the regime $z_{\rm max} \gg 1$ the integral over $z$ in eq.(\ref{RedshiftDistn}) remains constant as $F_\nu$ and hence $z_{\rm max}$ change (because $z_{\rm max}$ remains above the redshift over which the distribution falls to zero due to evolution).  Therefore, only the $F_\nu^{-\gamma}$ term in eq.(\ref{RedshiftDistn}) influences the source count distribution.  This argument is valid until $F_\nu$ increases to the the point at which $z_{\rm max}(F_\nu)$ has decreased sufficiently that the $z_{\rm max}$ cutoff strongly decreases the value of the integral over the redshift distribution (i.e. $z_{\rm max}$ falls below the redshift at which the term in square brackets in eq.(\ref{RedshiftDistn}) begins to change at high $z$).  When this occurs, there is a break in the source counts, and we enter the regime $z_{\rm max} \sim z_0$.

\medskip \noindent
iii) $z_{\rm max} \lesssim z_0$:  In this regime there is a break in the event rate count distribution for all those luminosity functions which are dominated by source counts at high redshifts, $\gamma < 2$.  The behaviour of the source counts is then determined by the integral of the bracketed terms in eq.\,(\ref{RedshiftDistn}) up to the limiting value of $z_{\rm max}$.  The break is caused by the fact that $z_{\rm max}$ decreases with increasing $F_\nu$ until the point at which it moves below the peak of the redshift distribution.  

Figure \ref{fig:xiplot} shows that for $\gamma<2$ the redshift distribution declines rapidly for redshifts below the peak, and the slope of this decline determines the slope of the number counts in the post-break region of the source count distribution.  Three parameters influence the slope of the redshift distribution: (a) the more sharply peaked the redshift evolution function $\psi(z)$, the steeper the source counts, (b) the flatter the luminosity function, the steeper the source counts and (c) the more negative $\alpha$, the steeper the source counts.  

Analysis of the behaviour of the fluence counts is particularly simple in the limit $z_{\rm max} \lesssim 1$.  Here, using the fact that $D_M \sim c z/H_0$ to a good approximation for $z \lesssim 1$, the cutoff limit is $z_{\rm max} \approx (H_0/c) \sqrt{E_{\rm max}/4 \pi F_\nu} \equiv C F_\nu^{-1/2}$.  The fluence counts therefore scale as
\begin{eqnarray}
\frac{d{\cal R}}{dF_\nu} \propto F_\nu^{-\gamma} \int_0^{C F_\nu^{-1/2}} z'^{4-2 \gamma} (1+z')^{\alpha(\gamma-1)-1} \frac{\psi(z')}{E(z')}\,dz' . 
\end{eqnarray}
In the limit $z_{\rm max} \ll 1$, we can ignore the evolution of the terms involving $(1+z)$, $\psi(z')$ and $E(z')$, which are all slow compared to the $z^{4-2 \gamma}$ term.  We thus have $d{\cal R}/dF_\nu \propto  F_\nu^{-\gamma} [z^{5-2 \gamma}]_0^{C F_\nu^{-1/2}} \propto F_\nu^{-5/2}$, and we see that the distribution tends to a Euclidean scaling.  

\section{Redshift and dispersion measure distributions} \label{sec:RateandDMcompare}

Although the event rate fluence distribution provides a conventional way to analyse the characteristics of an astrophysical population, in the case of FRBs the extra information provided by their DMs potentially enables us to investigate the redshift evolution of the phenomenon in a much more direct way.  In this section we examine the redshift distribution of events for a survey that integrates down to some limiting flux density or fluence. We then relate this to the more observationally-relevant DM distribution, under the assumption that the intergalactic medium contributes substantially to the DMs of FRBs.  The simplicity of this approach obviates the need to perform any integration over the evolutionary history, $\psi(z)$, enabling direct comparison with the DM distribution of FRBs found in a fluence- (or flux density-) limited survey.

\begin{figure}
 \begin{tabular}{cc}
 \epsfig{file=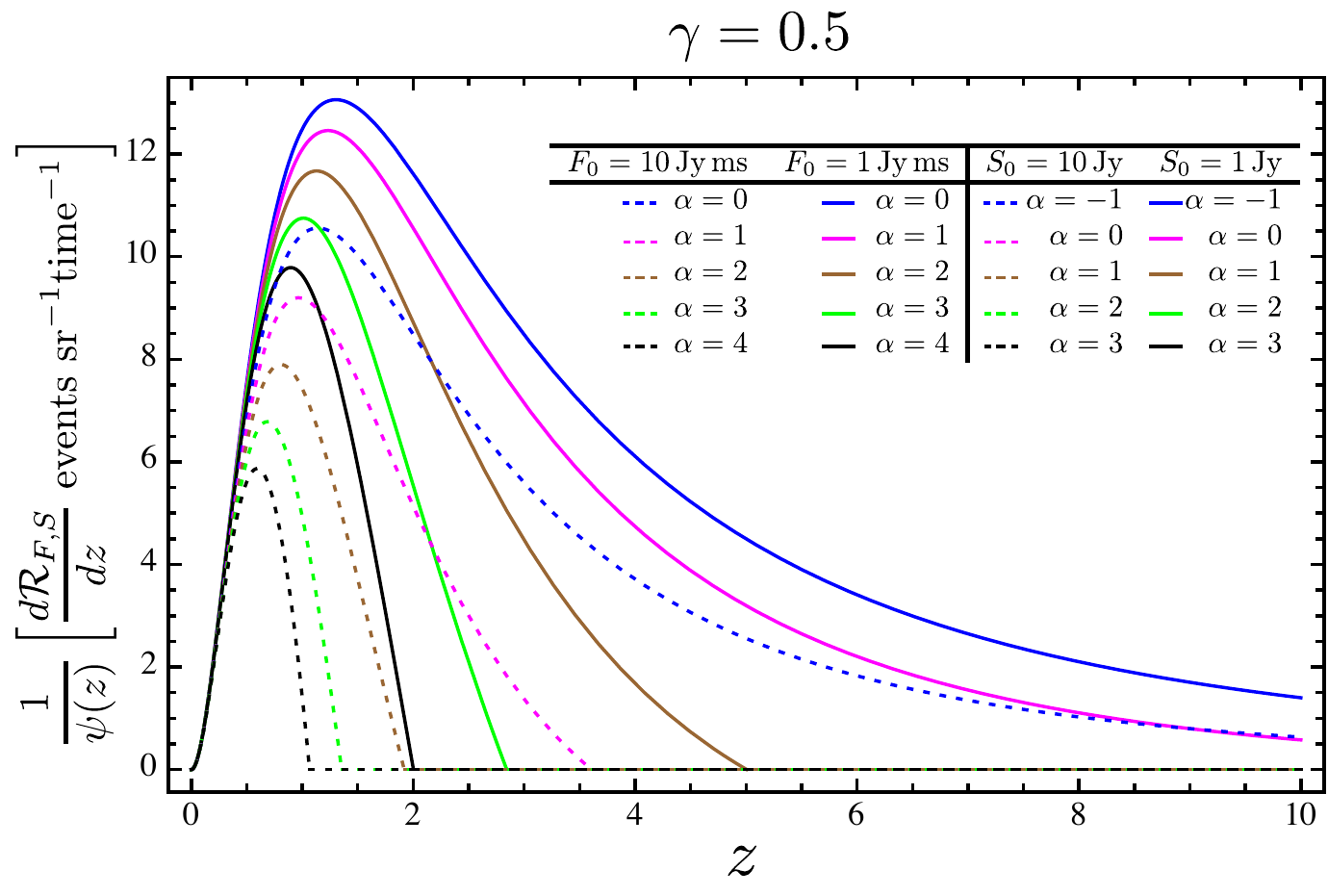,scale=0.6} &  \epsfig{file=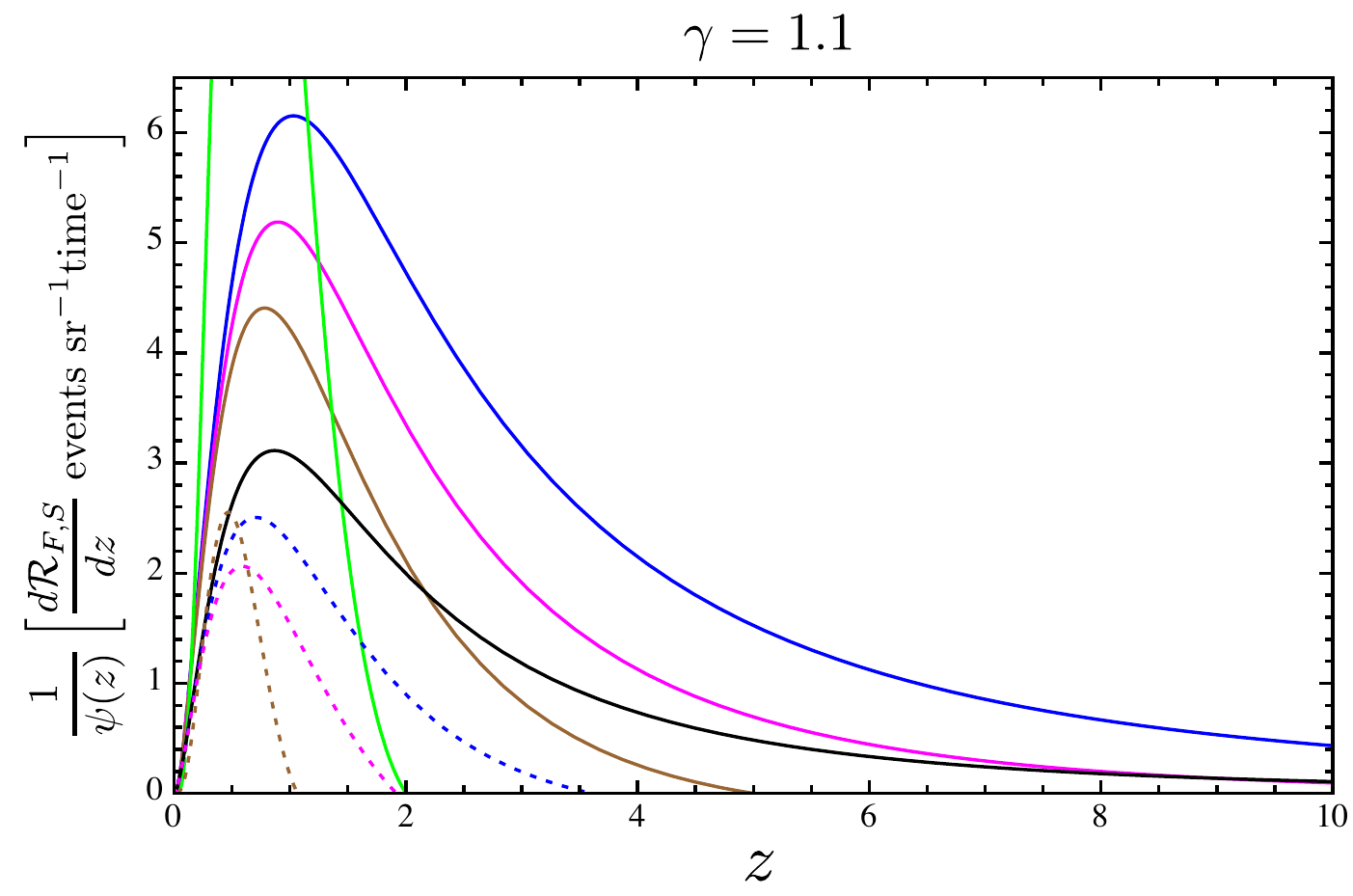,scale=0.6} \\
  \epsfig{file=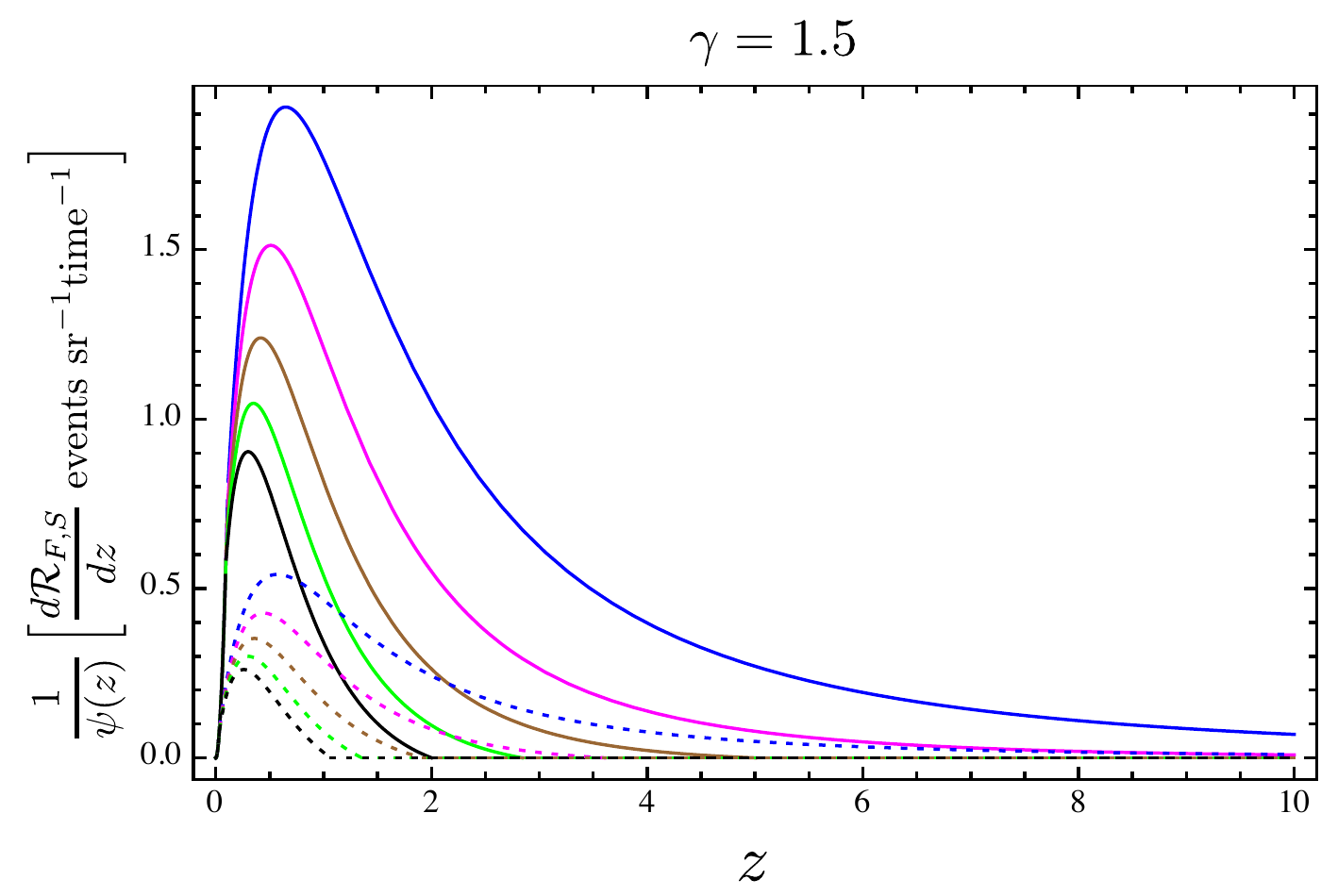,scale=0.6} &  \epsfig{file=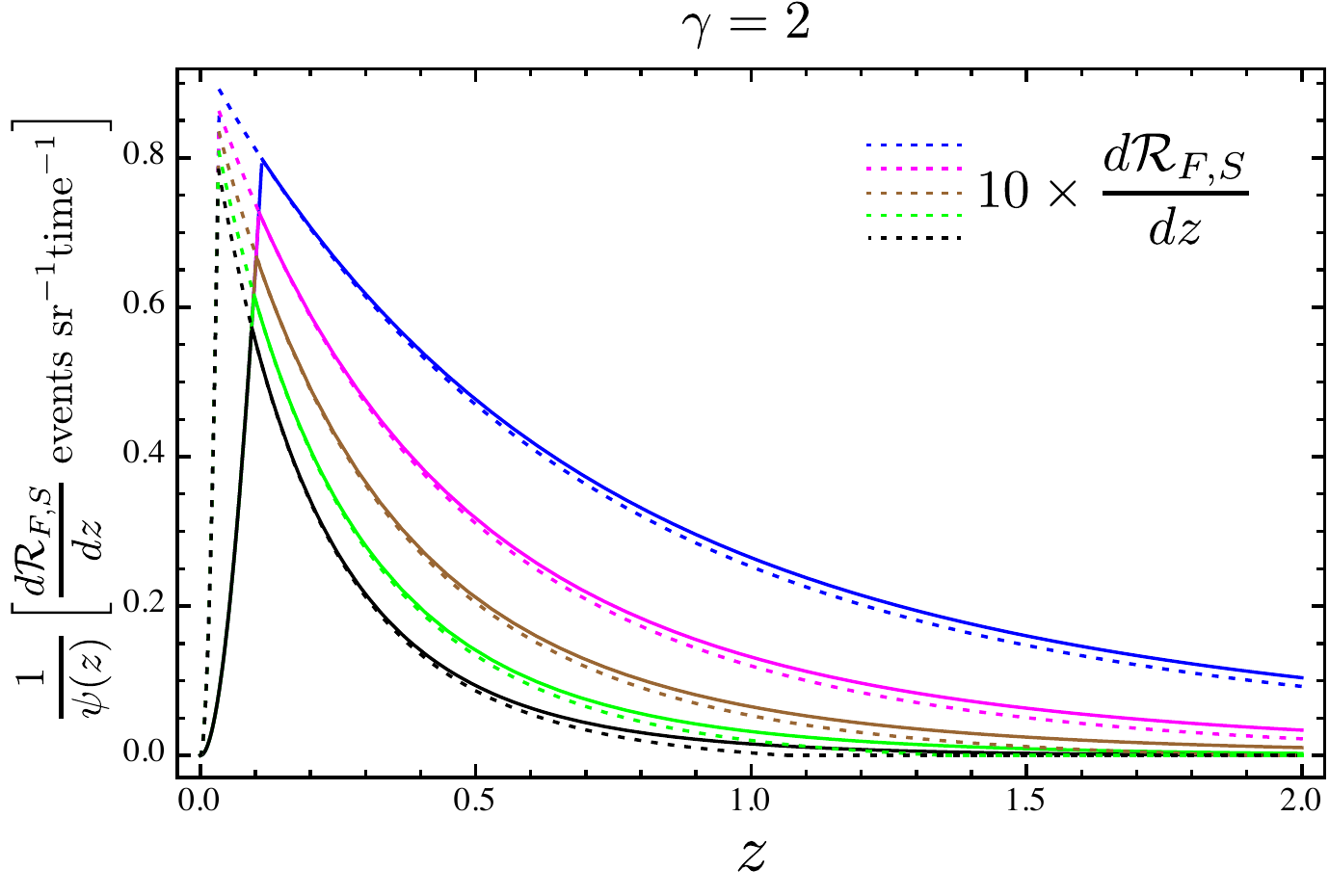,scale=0.6} \\
 \epsfig{file=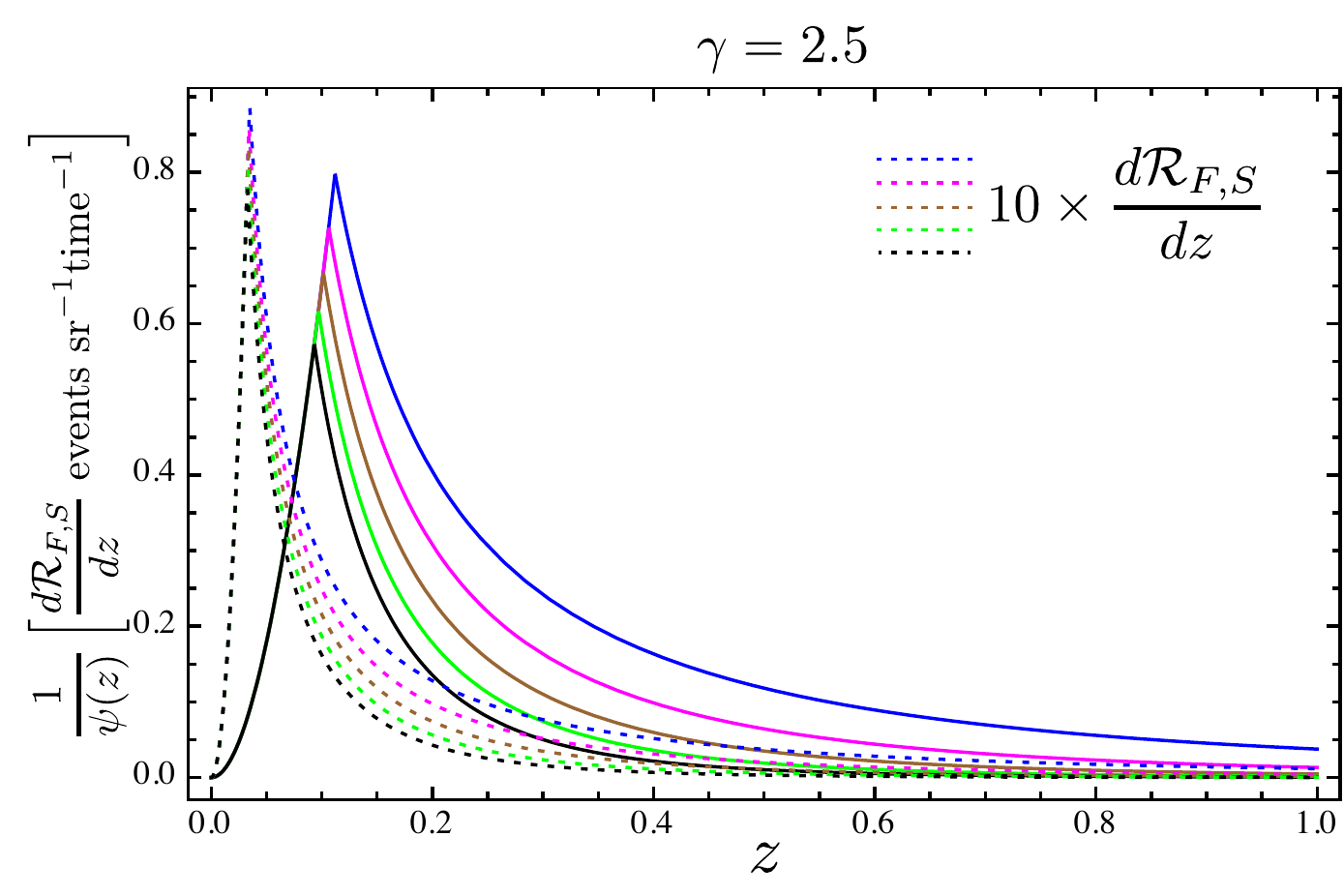,scale=0.6} &  \epsfig{file=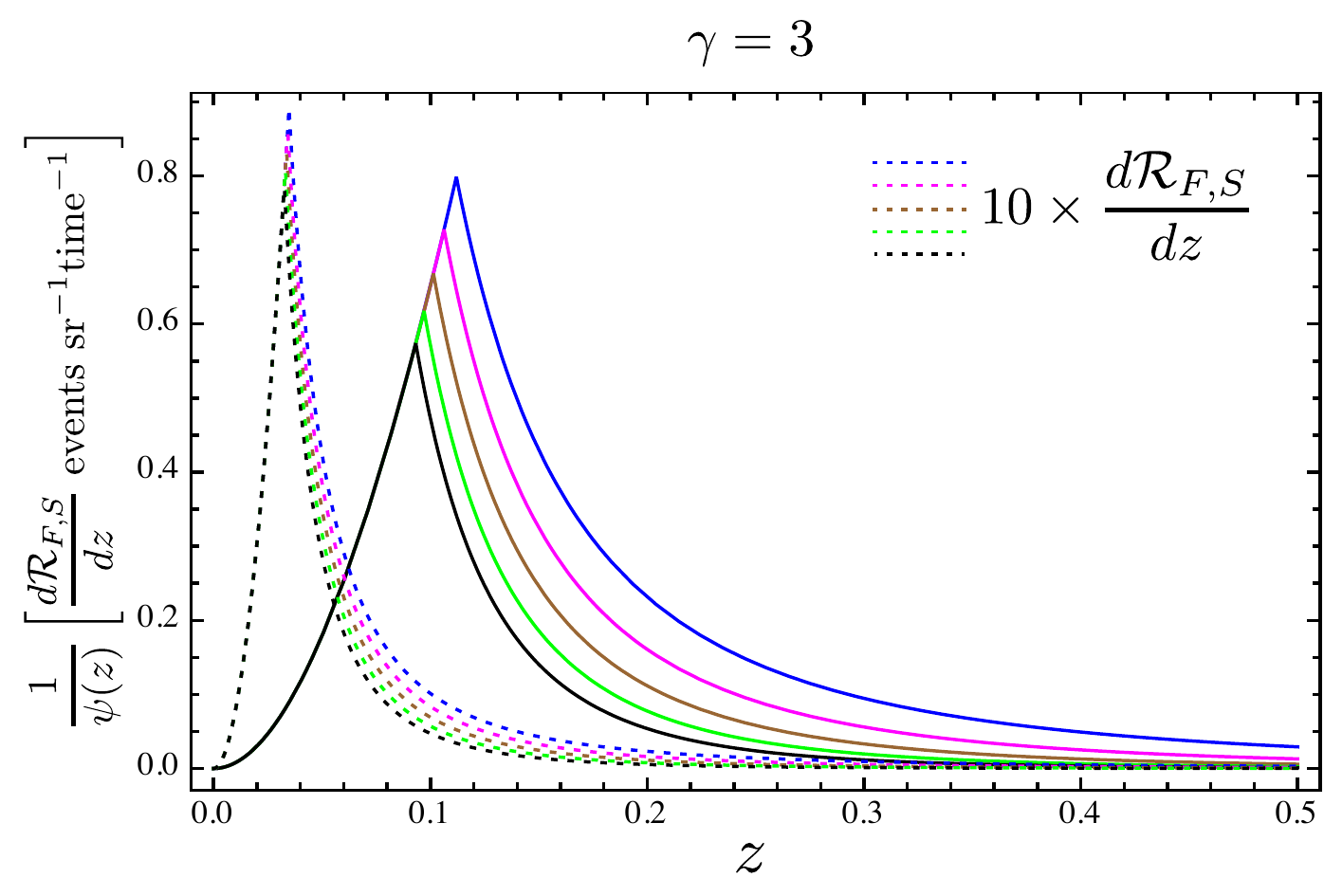,scale=0.6} \\
 \end{tabular}
 \caption{The redshift distributions for a survey that is limited in fluence to events exceeding $1\,$Jy\,ms (solid lines) and $10\,$Jy\,ms (dotted lines), for a range of energy function slopes and spectral indices.   The energy function used here extends a factor of 100 both higher and lower than a fiducial energy density, set so that a burst with a spectrum $F_\nu \propto \nu^{-1}$ (i.e. $\alpha=1$) has a fluence of 1\,Jy\,ms at $z=1$.  The evolution of the FRB event rate density, $\psi(z)$ (expressed in events per unit time per Gpc$^3$ of volume) is normalised out of these curves, so that the distributions purely reflect the interplay between telescope sensitivity and the volume of space probed for a given luminosity function and spectral index. 
Given the close correspondence between fluence and flux density distributions, these curves equivalently represent the redshift distribution for a survey with a limiting flux density of $1\,$Jy (solid lines) or $10\,$Jy (dotted lines), normalised so that a burst with a flat spectrum in flux density has a flux density of 1\,Jy at $z=1$.  In mapping between fluence and flux density one decrements the value of $\alpha$ by 1 (i.e. $\alpha \rightarrow \alpha -1$).  The key in the top left panel indicates the link between curve colour and spectral index for both fluence- and flux density-limited surveys.   For $\gamma=2,2.5$ and 3, the dotted distributions are plotted at ten times the actual rate for legibility.
 } \label{fig:Zdistns}
 \end{figure}

We derived above the redshift distribution of events at a given flux density or fluence.  To find the redshift distribution of all events integrated down to some limiting flux density or fluence we integrate either eq.(\ref{SourceRate}) or eq.(\ref{SourceRateF2}) down to $S_0$ or $F_0$ respectively.  It is conceptually useful to define quantities $S_{\rm min}(z)$ and $S_{\rm max}(z)$ which correspond to the flux densities that events of luminosities $L_{\rm min}$ and $L_{\rm max}$ respectively would possess at the redshift $z$.  Specifically, the redshift distribution for the power law luminosity function is
\begin{eqnarray}
\frac{d{\cal R}_{S}}{dz}(S_\nu>S_0;z) &=& 4 \pi D_H^5 \left( \frac{D_M}{D_H} \right)^4 \frac{(1+z)^{\alpha}}{E(z)} \psi(z) \int_{S_0}^\infty \theta_L \left( \frac{4 \pi D_L^2}{(1+z)^{1-\alpha}} S_\nu' \right) dS_\nu'     \nonumber \\
&=& 4 \pi D_H^5 \left( \frac{D_M}{D_H} \right)^4 \frac{(1+z)^{\alpha}}{E(z)} \psi(z) 
\begin{cases}
0, & S_0 > S_{\rm max}, \\
\frac{(1+z)^{1-\alpha}}{4 \pi D_L^2} \left( \frac{S_{\rm max}^{1-\gamma} - S_0^{1-\gamma}}{S_{\rm max}^{1-\gamma} - S_{\rm min}^{1-\gamma} } \right),  & S_{\rm min} < S_0 < S_{\rm max}, \\[1.5mm]
\frac{(1+z)^{1-\alpha}}{4 \pi D_L^2}, & S_0 < S_{\rm min}. \\[1.5mm]
\end{cases}  
\end{eqnarray}
Similarly, using eq.(\ref{SourceRateF2}) and defining $F_{\rm min}(z)$ and $F_{\rm max}(z)$ as the fluences that correspond to events of energies $E_{\rm min}$ and $E_{\rm max}$ at the redshifts, $z$, and integrating over a power-law energy function, we obtain the fluence-limited redshift distribution
\begin{eqnarray}
\frac{d{\cal R}_{F}}{dz}(F_\nu>F_0;z) &=& 4 \pi D_H^5 \left( \frac{D_M}{D_H} \right)^4 \frac{(1+z)^{\alpha-1}}{E(z)} \psi(z)
\begin{cases}
0, & F_0 > F_{\rm max}, \\
\frac{(1+z)^{2-\alpha}}{4 \pi D_L^2} \left( \frac{F_{\rm max}^{1-\gamma} - F_0^{1-\gamma}}{F_{\rm max}^{1-\gamma} - F_{\rm min}^{1-\gamma} } \right),  & F_{\rm min} < F_0 < F_{\rm max}, \\[1.5mm]
\frac{(1+z)^{2-\alpha}}{4 \pi D_L^2}, & F_0 < F_{\rm min}. \\[1.5mm]
\end{cases}  
\end{eqnarray}
Plots of the redshift distributions for a flux density or fluence-limited sample are shown in Fig.\,\ref{fig:Zdistns}. 
For a given redshift, a survey detects: (i) no events at that redshift if $S_0$ exceeds $S_{\rm max}(z)$ (ii) some fraction of events if $S_\nu$ falls between $S_{\rm max}(z)$ and $S_{\rm min}(z)$ or (iii) all events if $S_0$ falls below $S_{\rm min}(z)$.  In other words, the survey is complete to FRBs of a given flux density at some given redshift $z$ if $S_0 < S_{\rm min}$, but at some higher redshift it becomes only partially complete to events between $S_{\rm min}(z)$ and $S_{\rm max}(z)$. Eventually, at some yet-larger redshift, the value of $S_{\rm max}$ becomes sufficiently small that $S_{\rm max} < S_0$, and the survey detects no events beyond this point.
 
The dispersion measure (DM) distribution depends on the relation between redshift and dispersion measure.  In simple models of the IGM and the DM contribution from a host galaxy, there is a simple one-to-one mapping between the two quantities (Inoue 2003, Ioka 2004), and the DM distribution is a direct measure of the underlying redshift distribution of events.  In general, however, the IGM is not a homogeneous medium, there may be considerable dispersion in the DM values for an event at some redshift $z$, due to the intersection of the line of sight with the haloes of a discrete number of galaxies, as examined by e.g. McQuinn (2014).  We define $p({\rm DM}|z)$ as the DM probability density of events at some given value of $z$. Thus, in general, the resulting DM distributions for a flux density or fluence-limited samples are
\begin{eqnarray}
\left( \begin{array}{c}
\frac{d{\cal R}_S}{d{\rm DM}} (S_\nu>S_0) \\[1.5mm] 
\frac{d{\cal R}_F}{d{\rm DM}} (F_\nu>F_0)
\end{array} \right) 
&=& \int_0^\infty dz \,  \left( \begin{array}{c} 
\frac{d{\cal R}_{S}}{dz} \\[1.5mm] 
\frac{d{\cal R}_{F}}{dz}
\end{array} \right) p({\rm DM}|z). 
\end{eqnarray} \label{DMdistn}

It is beyond the scope of the present paper to investigate the DM histograms of models which take into account baryonic feedback and clumpiness in the IGM density. Although our formalism is sufficiently general to incorporate it, we defer a full treatment of feedback and IGM inhomogeneity, in which the shape of the distribution $p({\rm DM}|z)$ varies strongly with redshift, to a future work. Since our present purpose is to elucidate the essential features of the DM histogram, we adopt the simple model in which the IGM is treated as homogeneous, so that the DM distribution is represented by $p({\rm DM}|z) = \delta ({\rm DM} -\overline{\rm DM}(z))$, where the expectation of the dispersion measure at a given redshift is
\begin{eqnarray}
\overline{\rm DM}(z) = \frac{3  H_0 c}{8 \pi G m_p} \Omega_b \int_0^z \frac{(1+z) f_e(z)}{\sqrt{(1+z)^3 \Omega_m + \Omega_\Lambda}} dz'. \label{HomogIGM}
\end{eqnarray}
where $f_e(z) = \frac{3}{4} X_{e,H}(z) + \frac{1}{8} X_{e,He}(z)$ where $X_{e,H}$ and $X_{e,He})$ are the ionization fractions of Hydrogen and Helium.  We take $X_{e,H}=1$ for all $z<8$ and $X_{e,He}=1$ for $z<2.5$ and zero otherwise. We assume, for the sake of expediency that the epoch of Helium reionization occurs at $z=2.5$ in a sharp transition\footnote{There exists considerable uncertainty in the redshift range over which He reionization occurs, and the duration of this transition, but since the purpose of the current calculation is to demonstrate the fundamental appearance of such a phase transition in the DM histogram, we do not examine this point in detail. The epoch of He reionization is thought to occur in the range $3 < z < 4$ \citep{Sokasianetal02}, but recently detected variations in the effective optical depth of the He II Ly$\alpha$ forest at $z \gtrsim 2.7$ raise the possibility that the Universe is still undergoing He reionization at this epoch \citep{Daviesetal17}.}. In this case the flux density-limited DM histogram is
\begin{eqnarray}
\frac{d{\cal R}_S}{d{\rm DM}} ({\rm DM})= \frac{d{\cal R}_S}{dz}  (S_\nu > S_0; z_i) {\bigg /} \left. \frac{ \overline{\rm DM}(z)}{dz} \right\vert_{z=z_i} , \hbox{ where } \overline{\rm DM}(z_i) = {\rm DM},
\end{eqnarray}
and a relation identical in form applies for $d{\cal R}_F/d{\rm DM}$. 

\begin{figure}
 \begin{tabular}{cc}
 \epsfig{file=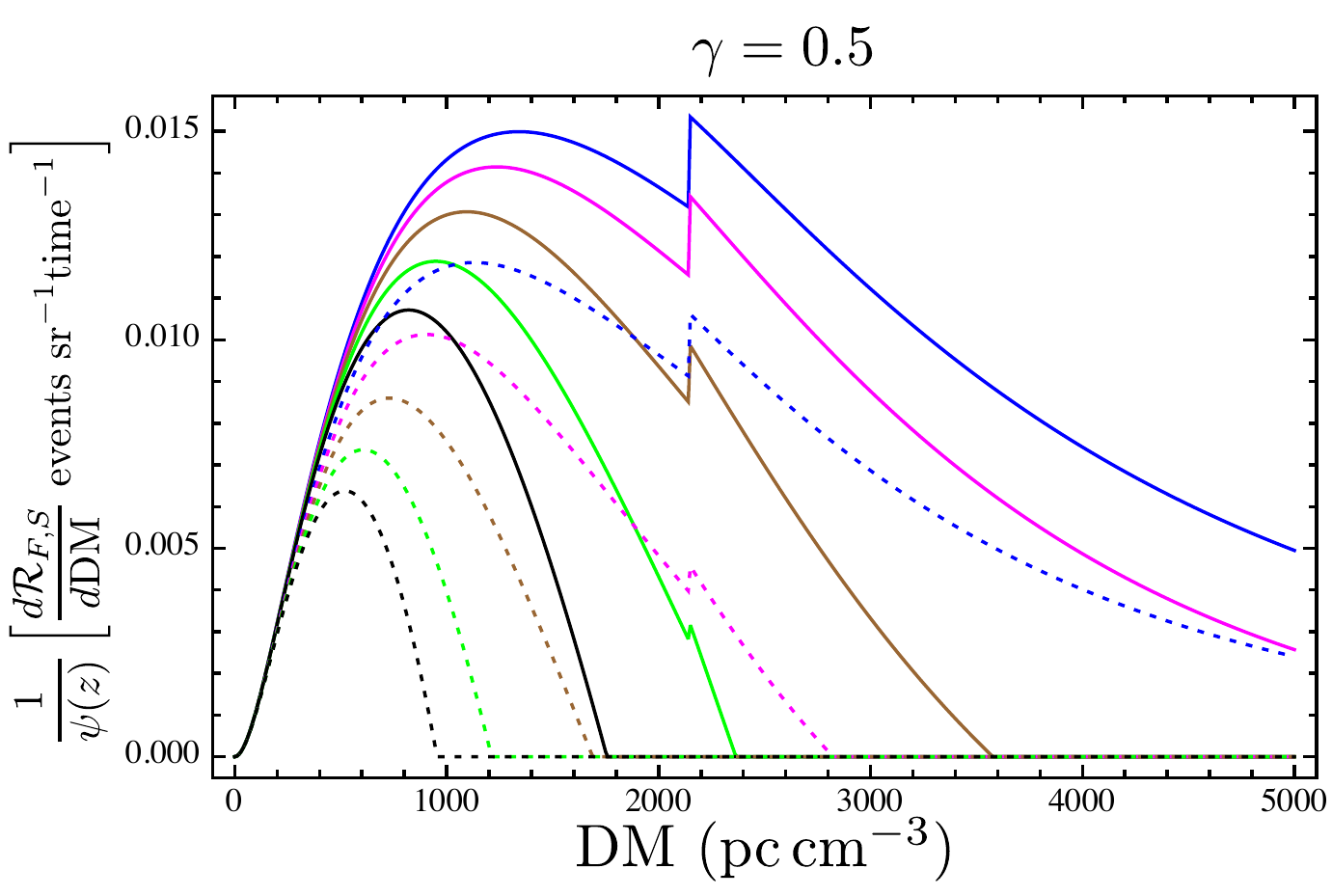,scale=0.6} &  \epsfig{file=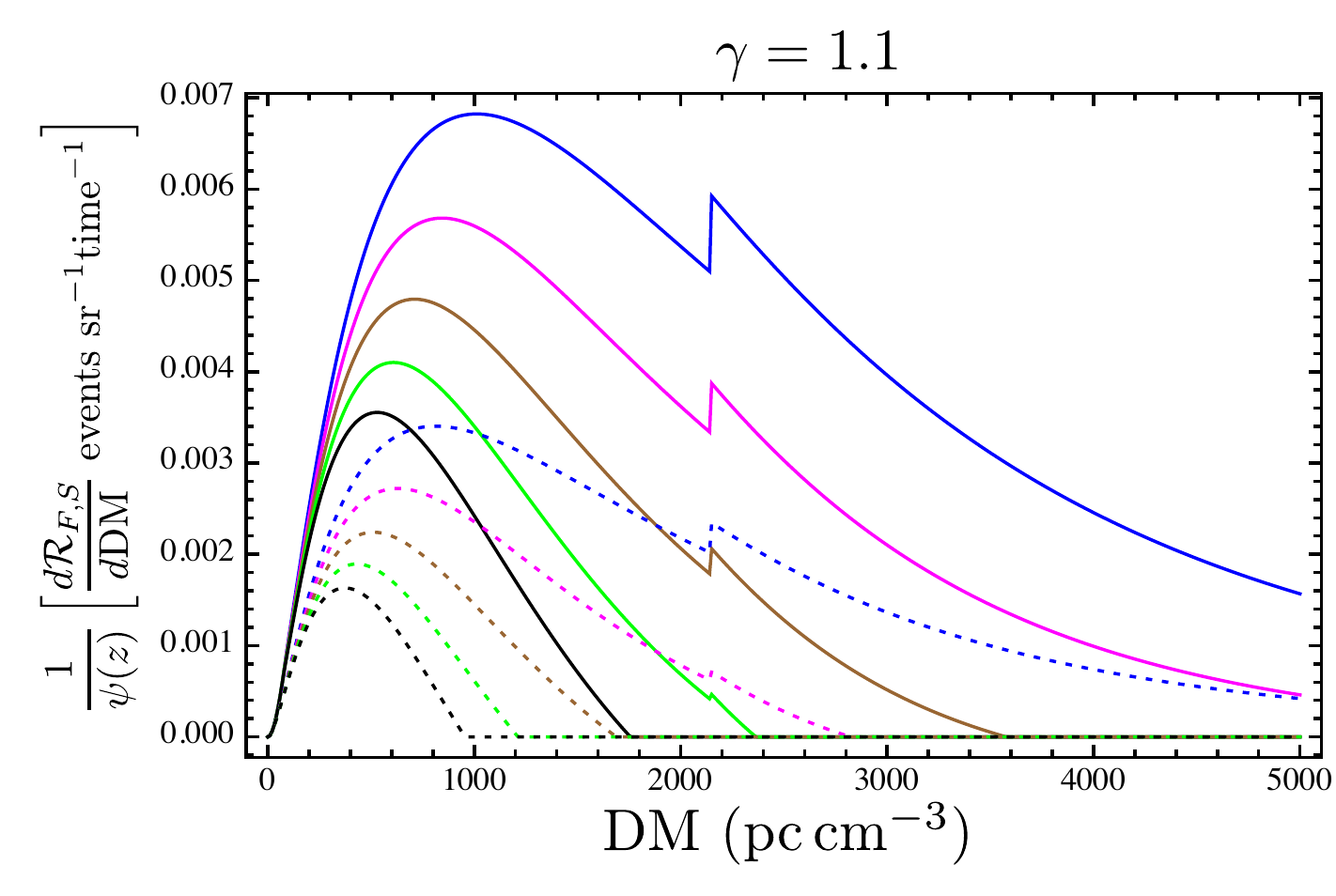,scale=0.6} \\
  \epsfig{file=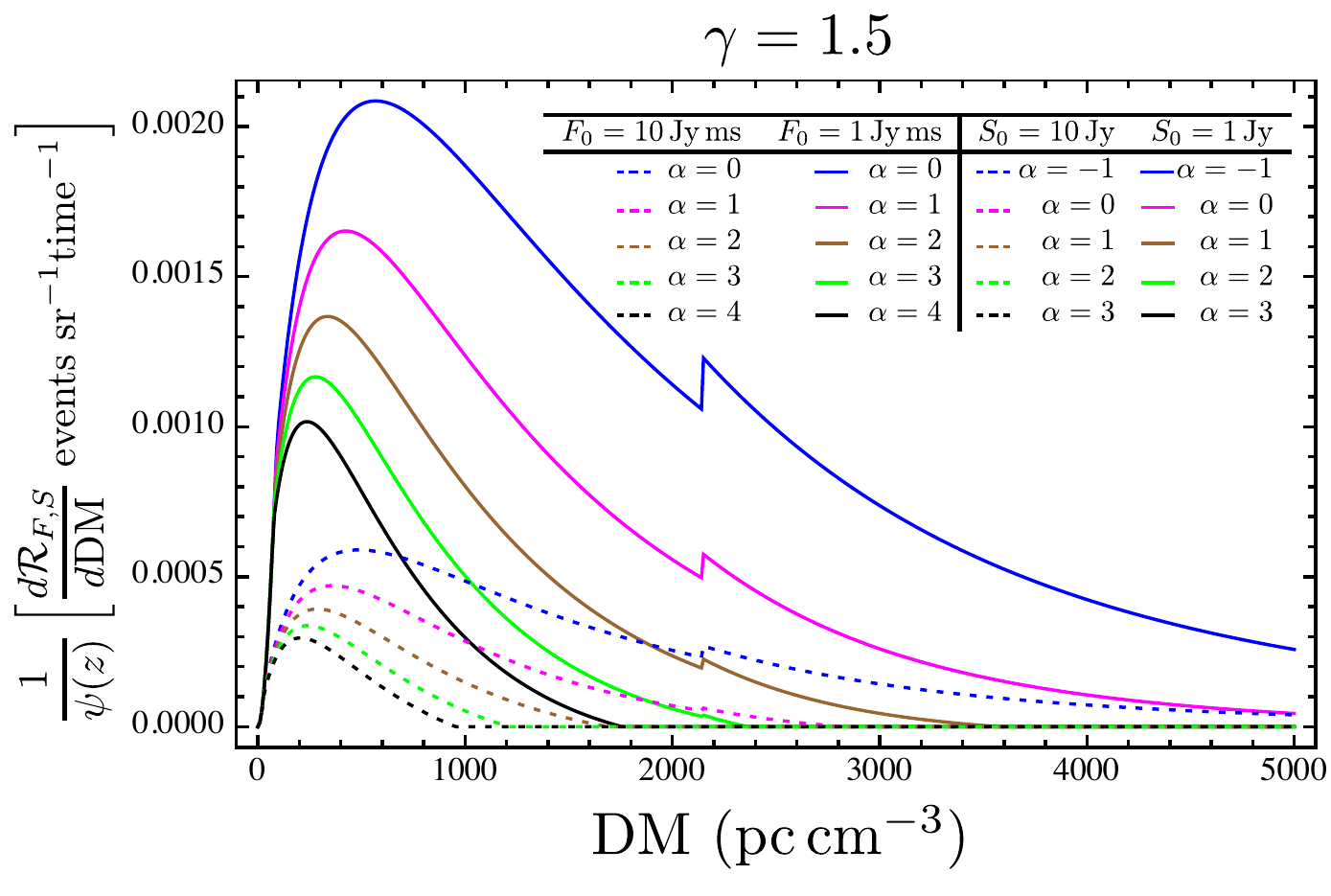,scale=0.6} &  \epsfig{file=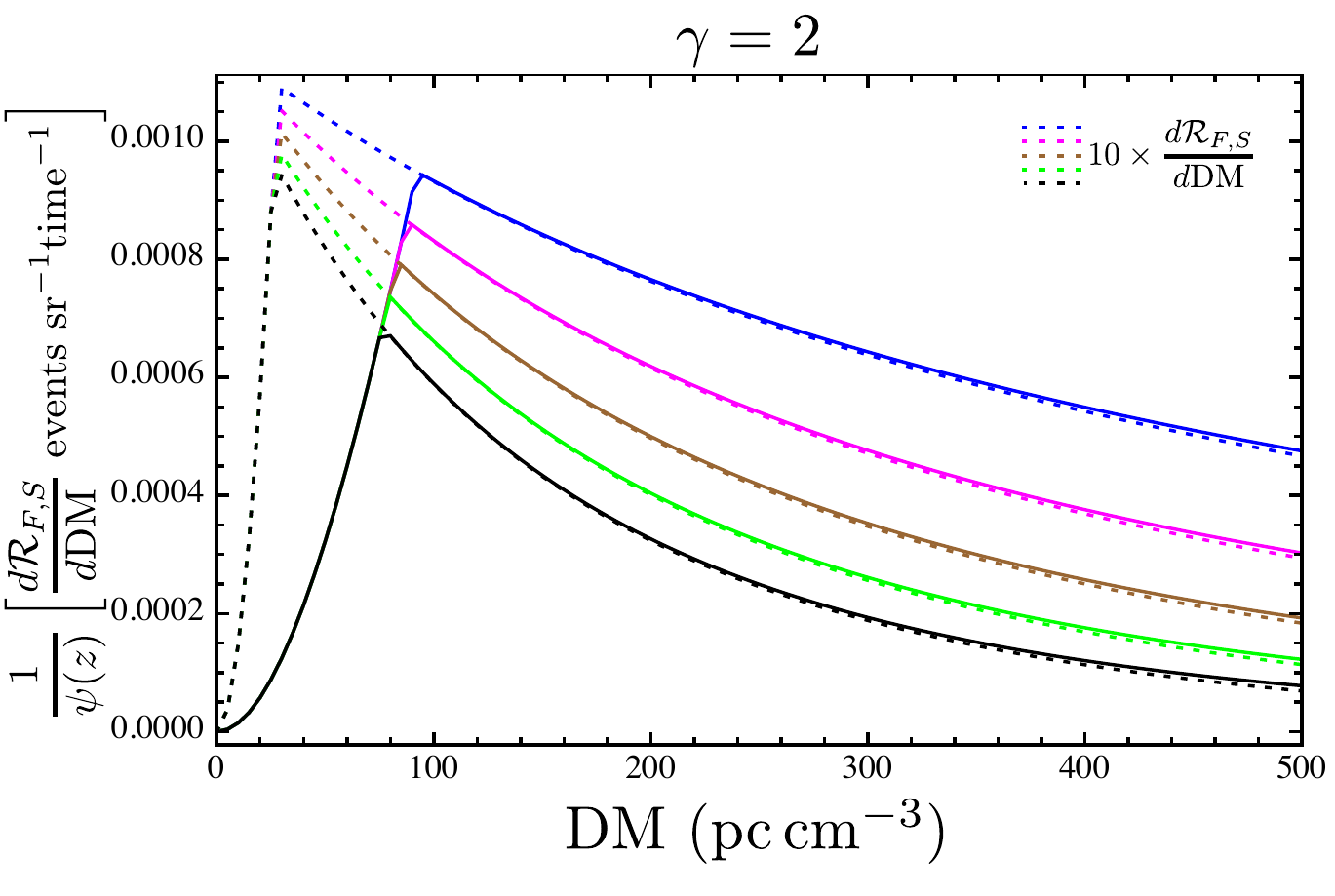,scale=0.6} \\
 \epsfig{file=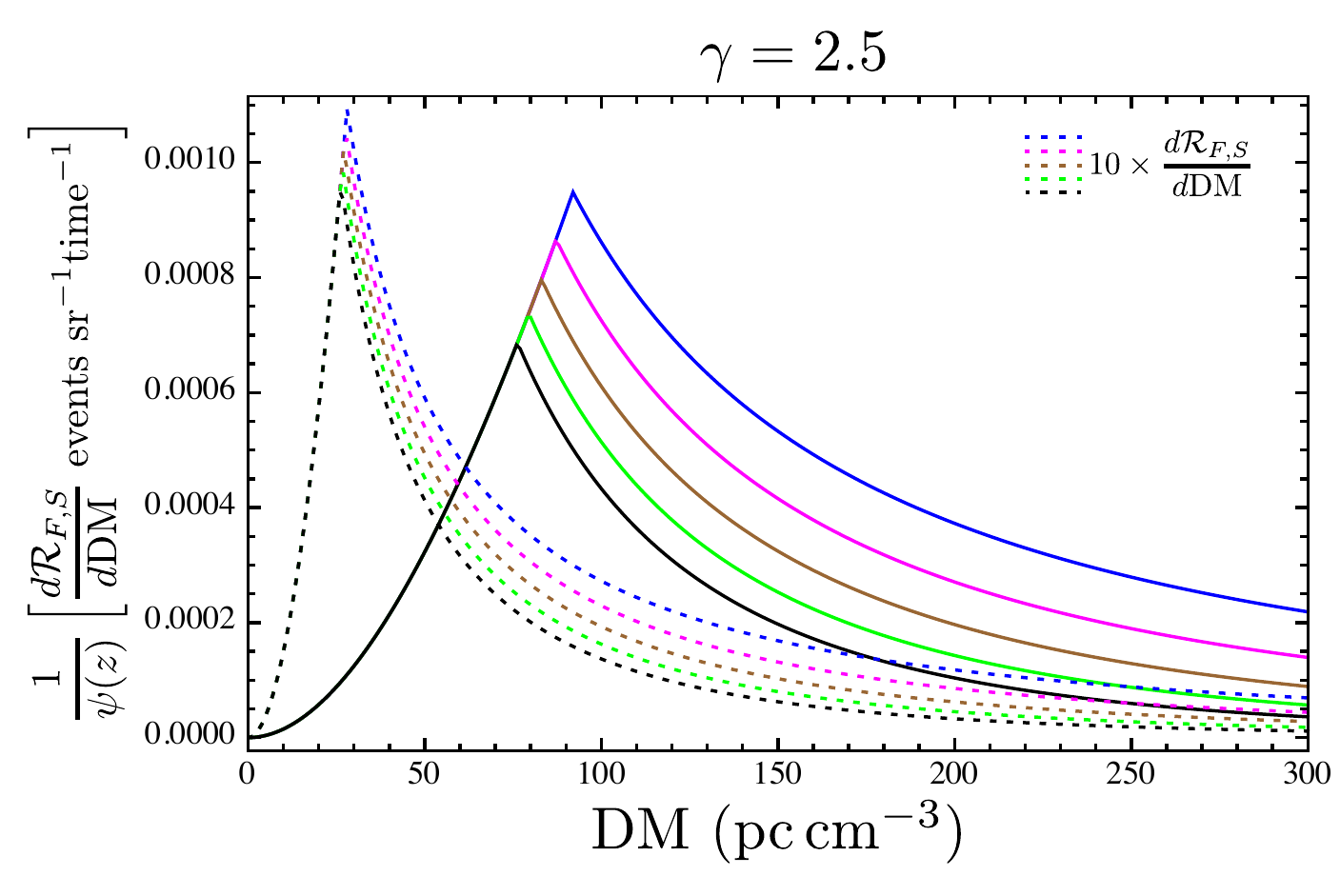,scale=0.6} &  \epsfig{file=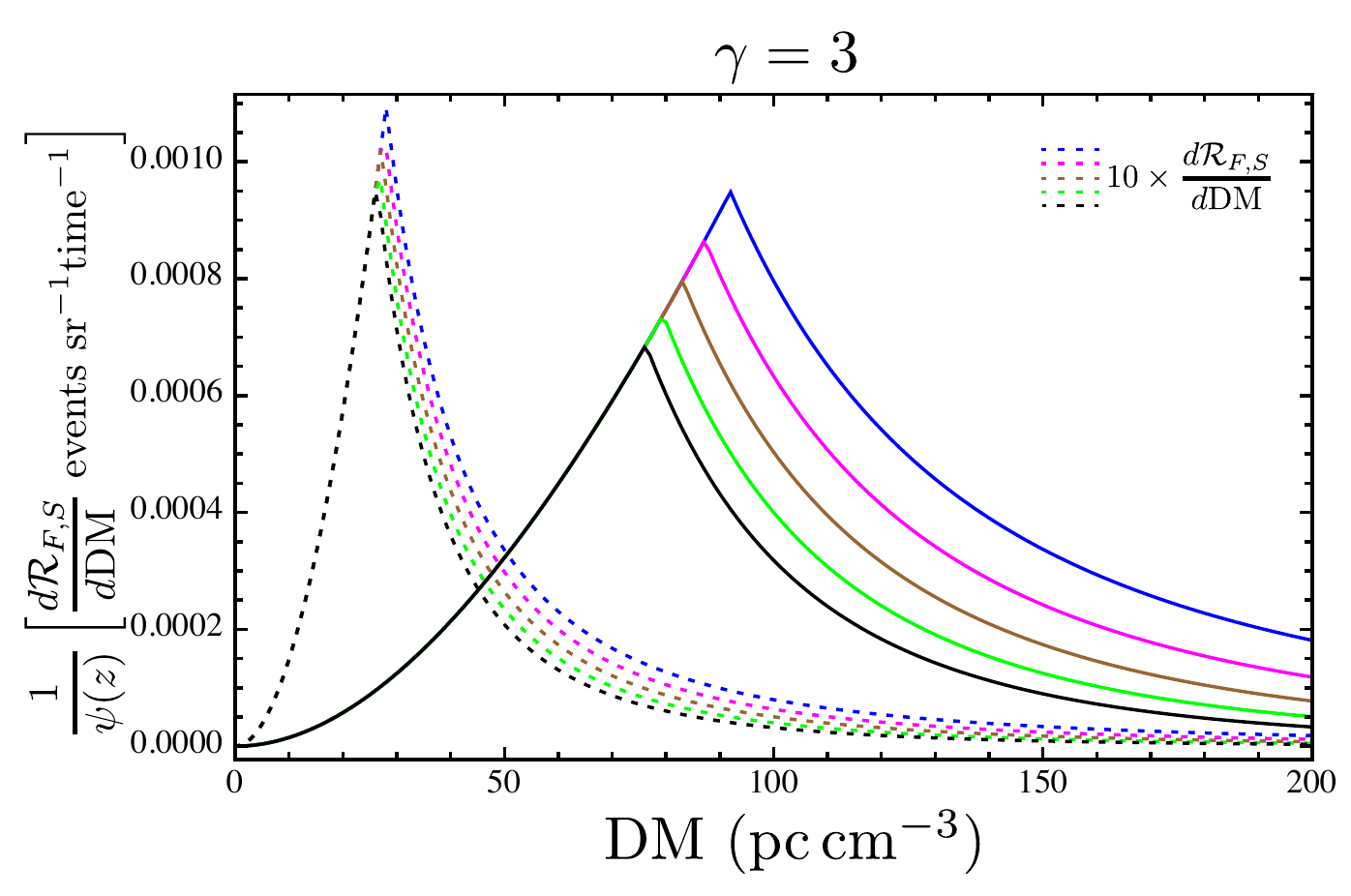,scale=0.6} \\
 \end{tabular}
 \caption{The dispersion measure distributions for flux density-limited survey redshift distributions shown in Figure \ref{fig:Zdistns}, using the mapping between $z$ and DM given by eq.\,(\ref{HomogIGM}).  Again, in mapping between fluence and flux density one decrements the value of $\alpha$ by 1 (i.e. $\alpha \rightarrow \alpha -1$), as shown in the legend.
 } \label{fig:DMdistns}
 \end{figure}

The resulting DM distribution functions are plotted in Fig.\,\ref{fig:DMdistns}.  A notable characteristic of these distributions is the transition in the qualitative behaviour at $\gamma = 2$.  We see that there is a strong bias towards the the detection of nearby, lower luminosity events in populations whose luminosity function is steeper than $\gamma=2$, and the DM distribution exhibits a cuspy peak at ${\rm DM} \lesssim 200\,$pc\,cm$^{-3}$ for the parameters chosen here.  Conversely, in models with flat luminosity distributions, $\gamma<2$, the distribution of observed events is biased toward the detection of more luminous events at greater distances, and the DM distribution extends to ${\rm DM}>2000-5000\,$pc\,cm$^{-3}$, with flatter (or inverted) spectra increasing yet further the range of detectable events.

We stress that the present treatment, and Figure\,\ref{fig:DMdistns} in particular, examines only the distribution of dispersion measures related directly to the IGM contribution which will include intervening clusters along the line of sight.  In practice, FRBs will incorporate DM contributions both from the Milky Way and from the host galaxy, including possible contributions from its cluster environment.  Comparison with the foregoing models is facilitated by either pre-subtracting the estimated host and Milky Way contributions of individual bursts, or by directly incorporating a zero-redshift offset directly into eq.(\ref{HomogIGM}) (i.e. by adding a constant to its RHS so that $\overline{\rm DM}(z=0) = {\rm DM}_0$).  Three further remarks on this point are in order. (i) One does not measure the electron column directly, but rather infers it from the fit to the time of arrival versus frequency; identical columns of plasma at two different redshifts along the line of sight contribute unequally to the time delay of the signal and hence the inferred DM. (ii) The contribution from the host galaxy relative to the IGM diminishes with increasing redshift. The electron column due to the IGM increases steadily with redshift since the baryon density in the Hubble flow increases as $(1+z)^3$ (e.g. at $z \approx 1$ the DM contribution from the IGM is $\approx 1000\,$pc\,cm$^{-3}$), whilst a fixed electron column DM$_0$ in a host galaxy produces a time delay that diminishes proportional to DM$_0 (1+z)^{-1}$ for a galaxy at redshift $z$. (iii) It is possible to solve for the mean the value of DM$_0$, the host galaxy contribution, in a sufficiently large sample of events by either (a) examining the minimum limiting value of DM in the sample, (b) regarding ${\rm DM}_0$ as a zero-point offset in the DM-$z$ relation and solving for it directly once the redshifts of some FRBs are known.   

\section{Discussion} \label{sec:Discussion}

 \begin{figure}
\centerline{ \epsfig{file=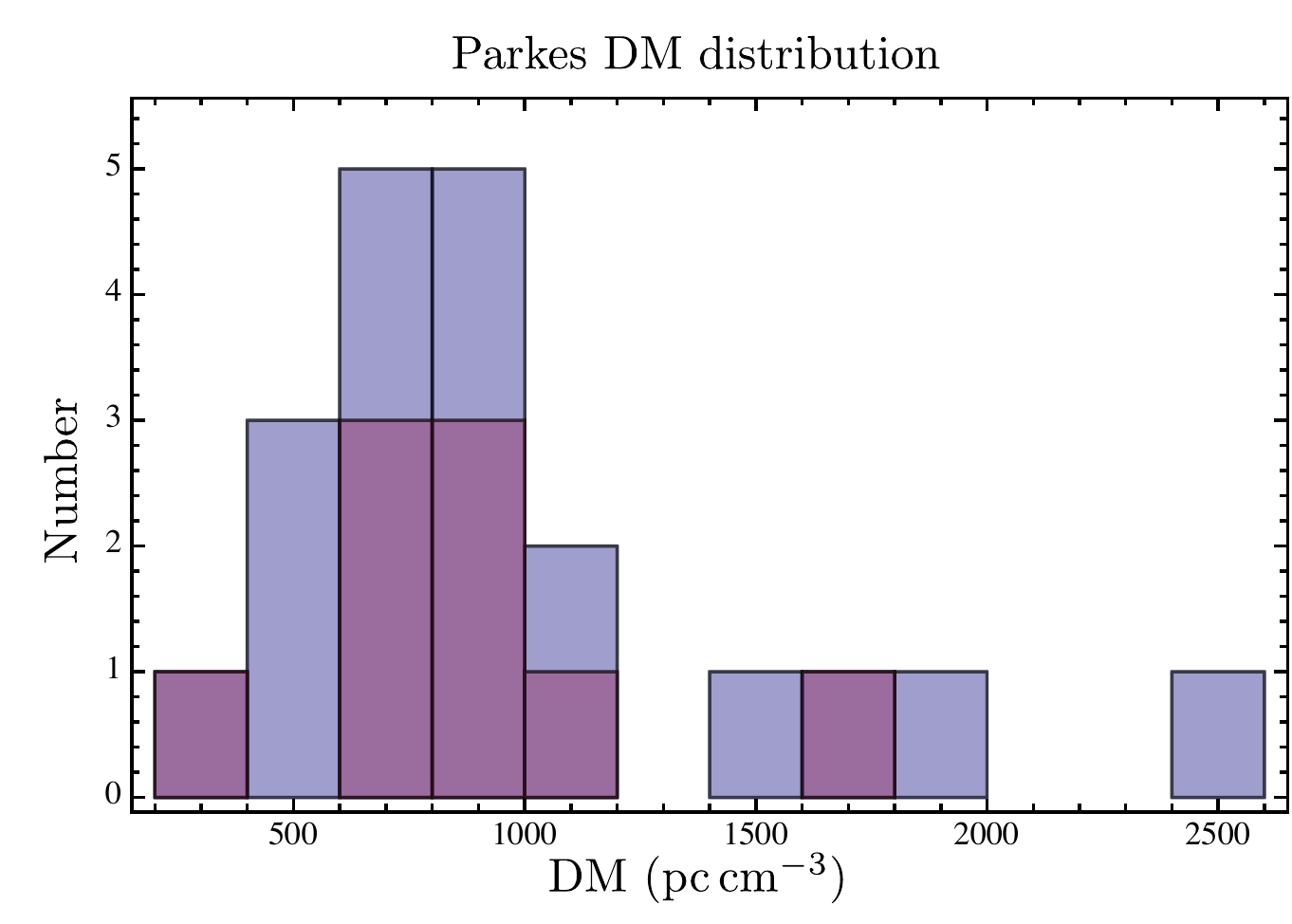,scale=0.6}} 
 \caption{The measured histogram DM of FRBs detected at Parkes, as listed in FRBCAT \citep{Petroffetal2016}, excluding the Lorimer burst (see Paper I). The blue histogram represents the distribution of the entire Parkes FRB sample, while the purple bars represent the nine events that exceed the Parkes 2\,Jy\,ms completeness fluence.
 } \label{fig:ObsDMdistn}
 \end{figure}

Despite the measurement of FRB event statistics being presently in their infancy, it is worthwhile attempting a qualitative comparison of the foregoing results against the basic observational properties of the population. The two obvious points of comparison are the DM distribution and the slope of the event rate fluence distribution.  Paper I showed that the maximum likelihood estimate of the integral rate distribution, ${\cal R}(F_\nu > 2\,{\rm Jy}\,{\rm ms}) \propto F_\nu^{\beta}$, of the FRB population detected with the Parkes telescope has an index of $\beta = -2.6_{-1.3}^{+0.7}$.  

Although existing limits on the counts slope are within two standard deviations of the Euclidean value of $\beta=-1.5$, it is useful to consider the implications of the steeper slope favoured by existing data.  In the context of an evolutionary model tied to the star formation rate we deduce that:
(1) A distribution that declines faster than Euclidean with increasing fluence only occurs near where the distribution turns over, at a fluence $F_\nu \approx E_{\rm max} (1+z_0)^{2-\alpha}/(4 \pi D_L(z_0)^2)$.  It is only evident for luminosity functions flatter than $\gamma \lesssim 2$.
(2) The steepness of the counts distribution persists only for an order of magnitude in fluence, before reverting to the Euclidean slope.
(3) The slope of the distribution in this range changes sharply, but the average slope, integrated over a factor of 10 in fluence past the peak in the distribution, decrements the slope by at most $\approx 1$, to an integral source counts slope of $\approx -2.5$ (i.e. $-3.5$ for the differential counts) for a population whose evolution follows the star formation rate.  The effects of evolution are more pronounced if the density scales quadratically with the star formation rate, in which case a decrement of up to $\approx 2$ is possible. Steeper slopes are only possible over a much narrower range in fluence.
(4) Steep slopes only occur for a population whose spectral index is $\alpha \lesssim 0$ for evolution linearly proportional to the star formation rate and $\alpha \lesssim 1$ for quadratic evolution.  The spectral indices of FRBs remains an open question at this stage.  Indeed, a more realistic scenario, although well beyond the level of sophistication merited by present data, would incorporate a distribution of spectral indices into the counts model.

The DM distribution of the FRBs detected at Parkes reported in FRBCAT \citep{Petroffetal2016}, is shown in Figure \ref{fig:ObsDMdistn}. We plot the distribution of both those events detected above the Parkes completeness fluence, and the entire Parkes FRB sample.  Two observational biases render the distribution additionally incomplete at high DM values: (i) dispersion smearing and (ii) temporal smearing caused by multipath propagation \citep[e.g. as discussed in][]{MacquartKoay2013}. In (i) the finite size of the spectral channels of the Parkes filterbank results in loss of sensitivity when the DM is sufficiently large that the dispersion smearing of the signal across an individual spectral channel exceeds the larger of the pulse width or the channel time resolution.  For the SUPERB Parkes observations \citep{Keaneetal2018}, this critical value is 1070\,pc\,cm$^{-3}$ at 1.4\,GHz for a 1\,ms pulse.  Although the S/N reduction due to dispersion smearing is quantifiable, the resultant change in the distribution is only correctable if the DM distribution as a function of limiting fluence is already known {\it a priori}.  Effect (ii) results in sensitivity loss when the smearing timescale exceeds the intrinsic burst duration (and detector temporal resolution).  The influence of temporal smearing on the DM distribution is harder to quantify, since the relation between DM and the smearing timescale, if one exists, is not established.  We note that the error and bias in the fluence of the Parkes events due to their unknown location within the beam does not affect the DM distribution.  

To the extent that only nine events comprise a fluence-complete sample, it is difficult to draw strong conclusions about the distribution of burst DMs.
We tentatively conclude that the existence of events at ${\rm DM} \gtrsim 1000\,$pc\,cm$^{-3}$, despite observational biases against their detection, disfavours models in which $\gamma \geq 2$, and points to a population of events whose luminosity distribution is shallow.  This conclusion is corroborated by the non-Euclidean index of the rate counts slope; it was shown in \S\ref{sec:powerlawtreatment} that $\gamma>2$ distributions generically produce rate counts slopes equal to the Euclidean value at high fluences.  Reaching this conclusion does assume that the large observed DMs are not dominated by contributions from their hosts.

Another notable point is that the shape of the DM distribution becomes increasingly insensitive to survey sensitivity as the luminosity distribution flattens.  For flat luminosity functions one is essentially able to detect events over the full range of redshifts over which the population exists, no matter what the survey sensitivity.  For instance, for a $\gamma = 0.5$ distribution and $\alpha=0$, we see that the detection rate at DM$=2000\,$pc\,cm$^{-3}$ for a survey with 10 times poorer sensitivity is $\sim 70$\% of the high-sensitivity detection rate. However, by $\gamma=2$, the corresponding detection fraction is $\sim 30$\%. For distributions $\gamma \geq 2$, the shape of the DM distribution changes qualitatively, and there is a strong dependence on the mean redshift with survey sensitivity.  This is because the slope of $\gamma \geq 2$ distributions strongly biases detection of events to the nearby Universe.  

Figures \ref{fig:Zdistns} \& \ref{fig:DMdistns} may be taken as predictions of how the redshift and DM distributions change as one goes from less sensitive surveys (e.g.\,with ASKAP \& UTMOST) to more sensitive surveys, such as those conducted by Parkes.  Results from these surveys will already constrain the value of $\gamma$, and that will allow us to infer the evolution function from the counts.  

The reality of a $\gamma < 2$ luminosity distribution would bear a number of consequences for the detectability of FRBs.  Although a rate counts distribution that is steeper than Euclidean at the high fluence end nets events at a relatively lower rate, paradoxically sensitivity is not the primary determinant in the detection of high-DM (high redshift) events.  This is analogous to the case for Active Galactic Nuclei, for which the luminosity function is sufficiently shallow over a large range in luminosity that it is possible to detect bright examples at high redshift \citep[e.g.][]{Wall83}.

Any cosmological features imprinted in the DM distribution, such as that due to a sharp He reionization signature, are more readily detectable with a flatter luminosity function and a flatter (or inverted) spectrum.   An important consequence of this is that one does not strictly require the redshifts of FRBs (or, more correctly, their host galaxies) to be able to undertake cosmology with FRBs.  



\section{Conclusions} \label{sec:Conclusions}

In this paper we have investigated the event rate distributions of the FRB population in terms of flux density and fluence, and redshift and dispersion measure.  We place particular emphasis on quantities involving the fluence because this quantity mitigates interpretational issues related to finite detector resolution and the effects of pulse temporal smearing due to multipath propagation. 

We summarise here the important points from each of these analyses, and detail their import for future studies of the population.

Even for events that emit as standard candles or standard batteries (i.e. with constant spectral energy density), the slope of the event rate distribution is a continuously changing function of fluence and, equivalently, flux density.  However,  the energy and luminosity distributions of the FRB population are likely to be broad.  If, after accounting for the contributions of the host galaxy and the Milky Way, we interpret the bulk of the DM as a measure of distance, and taking into account the range in burst flux densities, we would infer that FRB luminosities span a broad range.  We remark that a model incorporating a broad luminosity function can explain both the observed DM distribution and a large scatter in the DM-fluence relation, under the assumption that the IGM dominates the observed DM, without having to appeal to significant host DM contributions.  However, it presently remains an open question whether the IGM contribution does indeed dominate the total DM.

From studies of other cosmological populations such as AGN, it is known that for such a broad luminosity distribution the event fluence distribution separates into three distinct regions.  Normalised by the Euclidean source counts slope, $F_\nu^{-5/2}$ this then consists of a (i) low-fluence region that rises with fluence, (ii) a broad plateau and (iii) a high-fluence region in which the distribution declines with increasing fluence until, at very high fluence, it declines proportional to the Euclidean scaling of $F_\nu^{-5/2}$.

The high fluence end of the rate distribution, region (iii), is dominated by the distribution of the population on cosmological scales, and not by the luminosity function of the bursts. The width of the maximum between regions (ii) and (iii) depends more on the luminosity function than the cosmology.  For shallow energy (luminosity) distributions $E_\nu^{-\gamma}$ with $\gamma < 2$, there is a prominent break (evident as a maximum in the Euclidean normalised counts), which occurs at a fluence  $F_\nu \approx E_{\rm max} (1+z_0)^{2-\alpha}/(4 \pi D_L(z_0)^2)$ (or, equivalently, $S_\nu \approx  L_{\rm max} (1+z_0)^{1-\alpha}/(4 \pi D_L(z_0)^2)$ in the flux density distribution), where $z_0$ is a characteristic redshift beyond which the population density begins declining sharply.  By contrast, the breadth of the plateau at lower fluences, region (ii), is determined by the width of the energy (luminosity) distribution.

There is a large motivation to undertake large field-of-view surveys to characterise the high fluence tail of the FRB distribution. It is this region that encodes the most useful information about the cosmological evolution of the population; observations that probe below the turnover instead yield information on the luminosity distribution of the population.  This presents an important distinction between the statistics of FRBs and non-transient phenomena such as AGNs. The source counts statistics of static populations can only be improved in regions (i) and (ii), since generally the whole sky is already characterised at high flux densities.  However, continued observations of the bright FRB population are able to progressively refine knowledge of their evolutionary history.  

An important unresolved question is related to the existence or otherwise of a direct relationship between fluence and distance, for FRBs.  This depends on the slope of the energy (or luminosity) distribution of events; distributions with power law indices flatter than $\gamma=2$ contain a large fraction of observed events at large distances, and with steeper distributions the observations will be dominated instead by an overwhelming fraction of nearby events.  
This critical value of the power law of the luminosity function was first recognised in early analyses of the quasar and radio galaxy populations; at the critical value the increasing numbers of fainter sources exactly cancel the decreasing volume in which they can be seen \citep{vonHoerner1973}.  The fluence - distance relation will have a very large scatter for any luminosity function whose slope is near the critical value. This situation is nearly the case for radio galaxies and quasars, yielding no clear relation between flux density and distance in region (ii).  For variations in the slope of the luminosity function either side of this critical value there is either a direct statistical relation between fluence and distance, or an inverse relation in which on average fainter objects are more likely to be closer.  For extragalactic radio source surveys the brightest known sources are AGN at large distances while the faintest radio sources are more likely to be nearby starburst galaxies.  It is possible that a similar situation applies to the FRB population; an energy function shallower than $\gamma \approx 2$  would enable even widefield surveys with sufficient sensitivity to access region (ii) to detect events over a large range of redshifts over which the population is distributed.

Observational constraints on the slope of the event rate distribution place bounds on the slope of the FRB energy and luminosity function.  
In particular, it is just possible to explain the steep fluence distribution suggested by existing Parkes data, but requires contrived conditions.  We have investigated this in the context of a cosmological population tied to some power of the star formation rate.  Although the counts slope changes continuously past the peak of the distribution in these models, an average integral counts slope of $\beta \approx -2.5$ is possible over a decade in fluence, compatible with the value $\beta = -2.6_{-1.3}^{+0.7}$ derived from the Parkes FRB sample \citep{MacquartEkers18,Bhandarietal17} .  However, such a steep slope, in which strong evolution is inherent to the population, would be expected to persist only over a factor $\sim 10$ range in fluence past the peak of the distribution, and requires that the luminosity function be flatter than $\gamma \lesssim 2$.  For a population that evolves linearly with the star formation rate, only  bursts whose spectra are either flat or rising with frequency exhibit this behaviour.  However, bursts whose fluence declines with frequency also exhibit this behaviour if the population evolves faster with redshift, as exemplified by a scenario in which the density changes quadratically with the star formation rate, a scenario which also approximates the evolution of the AGN phenomenon.

The event rate counts should revert to Euclidean at yet higher fluences as the horizon of observable events draws inwards, and the effects of population evolution and spacetime curvature become negligible over the volume of detectability.  

The generic power-law behaviour of the high-fluence region of the distribution holds implications for the influence of scattering and lensing effects on the event rate distribution.  We examine the manner in which diffractive scintillation contributes to the behaviour of the FRB counts at high fluence. The exponential distribution of event amplifications which is normally expected from diffractive scintillation declines more sharply than the $F_\nu^{-5/2}$ behaviour intrinsic to the distribution, indicating that the limiting form of the source counts should be a power-law in nature.  
Qualitatively different models \citep[e.g. those based on caustics produced by plasma lenses,][]{Cordesetal17} may produce different observed source counts distributions in detail, but whether FRBs are likely to be observed very far from their de-magnified (``intrinsic'') fluences still depends on the slope of the intrinsic counts distribution relative to the slope in the tail of the magnification probability distribution.


The FRB dispersion measure distribution holds the potential to be both a new probe of cosmological physics, and a further measure of the burst energy and luminosity function, and the spectral index. The behaviour of the DM distribution depends on the slope of the FRB luminosity function.  As with the event rate distributions with fluence, there is a qualitative change in the character of the DM distribution at at $\gamma=2$, with flatter distributions probing to high redshifts, and with the peak of the distribution turning over slowly to higher DM values. By contrast, steeper distributions exhibit a sharp peak in the DM distribution, and the mean DM moves progressively lower as the the luminosity function steepens.  
  
For this reason the evolution of the DM distribution with survey sensitivity potentially yields further information on the underlying FRB luminosity and cosmological distributions. We remark that an analysis of this type could be readily undertaken by comparing the DM distributions of surveys with different sensitivities, such as those samples acquired by Parkes and ASKAP.

There is tentative evidence from the DM distribution of Parkes FRBs that favours a luminosity function flatter than $L_\nu^{-2}$.  If confirmed by a rigorous analysis of the Parkes DM histogram and its associated selection biases, this would confirm that sensitivity is not the primary determinant to detect distant bursts. We remark that in this context that the recent report of a DM 2600\,pc\,cm$^{-3}$ event \citep{Bhandarietal17} would therefore be unsurprising.

The DM distribution is potentially an extremely powerful probe of the baryonic content of the intergalactic medium and its cosmological development.   Baryonic feedback processes may play a large role in the interpretation of FRB DMs \citep{McQuinn2014}.  Another potential contributor to the shape of the DM distribution for events $z \gtrsim 2.5$ is He reionisation.  We have shown that the effect of this phase transition is in principle observable in the DM distribution.  However, the scenario we consider is optimistic: the DM signature is expected to be diluted if the transition is not abrupt. Random variations in the DM along individual lines of sight caused by feedback, resulting from inhomogeneity in the IGM destroying the direct relation between DM and $z$ along different sight lines, will further dilute this signal.

Although FRBs have been heralded as potentially revolutionary cosmological tools \citep[e.g.][and references therein]{McQuinn2014,Macquartetal15}, it is pertinent to remember that GRBs were once similarly touted as cosmological tools, but that they largely failed to fulfil this expectation.  This is because no independent distance indicator was available for each GRB, and the distribution of GRB redshifts was highly biased by the selection effects in the optical followup process.  Many bursts were not followed up, and there large selection biases inherent in the  GRB events for which the host galaxy redshift was obtained.  FRBs offer renewed help in this regard, since the detection of events and the measurement of their associated DMs is not subject to the same strong selection biases inherent to the optical followup of GRBs.

\section*{Acknowledgements}
Parts of this research were conducted by the Australian Research Council Centre of Excellence for All-sky Astrophysics (CAASTRO), through project number CE110001020. This research was also partly supported by the Australian Research Council through grant DP180100857.

\bibliographystyle{mnras}
\bibliography{references-FRBCounts2}

\bsp	
\label{lastpage}
\end{document}